%% file: IBMQuantum.tex
\documentclass[
reprint,
amsmath,
amssymb,
aps,
showkeys
]{revtex4-2}

\input{AbuGhanem}
\begin{document}

\title{IBM Quantum Computers: Evolution, Performance, and Future Directions}

\author{Muhammad AbuGhanem{$^{1,2}$}}
\address{$^{1}$ Faculty of Science, Ain Shams University, Cairo, 11566, Egypt}
\address{$^{2}$ Zewail City of Science, Technology and Innovation, Giza, 12678, Egypt}

\email{gaa1nem@gmail.com}

\date{\today}

\begin{abstract}

Quantum computers represent a transformative frontier in computational technology, promising exponential speedups beyond classical computing limits. IBM Quantum has led significant advancements in both hardware and software, providing access to quantum hardware via IBM Cloud\textsuperscript{®} since 2016, achieving a milestone with the world’s first accessible quantum computer. This article explores IBM’s quantum computing journey, focusing on the development of practical quantum computers. We summarize the evolution and advancements of IBM Quantum’s processors across generations, including their recent breakthrough surpassing the 1,000-qubit barrier. The paper reviews detailed performance metrics across various hardware, tracing their evolution over time and highlighting IBM Quantum’s transition from the noisy intermediate-scale quantum (NISQ) computing era towards fault-tolerant quantum computing capabilities.

\end{abstract}


\maketitle

\input{01Introduction}

\input{02IBM}

\input{03Hardware}

\input{04Performance}

\input{05Software}

\input{06Path}

\input{07Conclusions}

\input{08Acknowledgment}

\input{09References}
\input{10Appendix}

\end{document}

%% file: AbuGhanem.tex
\usepackage{graphicx}
\usepackage{dcolumn}
\usepackage{bm}
\usepackage{url}
\usepackage{hyperref}
\usepackage[braket, qm]{qcircuit}
\usepackage{graphicx}
\usepackage{float}
\usepackage{braket}
\usepackage{multirow}
\usepackage{mathrsfs}
\usepackage{hyperref}
\hypersetup{colorlinks,allcolors=blue,urlcolor=black}
\usepackage{enumitem}
\usepackage{xcolor}
\usepackage{tabularx}
\usepackage{color,soul}
\usepackage{wasysym}
\usepackage{tikz,lipsum,lmodern}
\usepackage[most]{tcolorbox}
\usepackage{xurl}
\usepackage{array}
\usepackage{xcolor} 
\usepackage{caption}

%% file: 01Introduction.tex
\section{Introduction}

Quantum computers represent a revolutionary approach to computation~\citep{NISQ24}, leveraging the principles of quantum mechanics~\citep{principles,AbuGMSc19} to potentially solve problems that are beyond the reach of classical computers~\citep{nisqQC11,nisqQC10,Light,PhotonicQuantumComputers}. The NISQ era~\citep{qsuperm1,NISQ18} marks a pivotal phase in quantum computing, characterized by rapid advancements and challenges in achieving practical quantum applications~\citep{NISQ24}.

Industry has been facilitating access to quantum computers for both academic and commercial users. Companies such as 
IBM (2016)~\citep{IntoAsnG53}, Rigetti Computing (2017)~\citep{IntoAsnG55}, IonQ (2020)~\citep{IntoAsnG91}, Honeywell (2020)~\citep{IntoAsnG90}, Google (2020)~\citep{IntoAsnG93,GoogleAI}, Xanadu  (2020)~\citep{IntoAsnG92}, Oxford Quantum Circuits OQC (2021)~\citep{IntoAsnG94}, PASQAL (2022)~\citep{IntoAsnG96}, QuEra (2022)~\citep{IntoAsnG95}, and Quandela (2022)~\citep{IntoAsnG97} have made their systems available via cloud platforms. While others have adopted a reseller model through web-based services~\citep{IntoAsnG91,IntoAsnG99,IntoAsnG98,IntoAsnG100}. As a result, research on quantum computers has increased significantly~\citep{IntoAsnG101}. For instance, scientific papers utilizing IBM's quantum systems via cloud service have reached approximately 2,800 as of February 2024, with over 3 trillion circuits executed through IBM Quantum platform~\citep{Assessing}.

IBM Quantum~\citep{IBMQ} has emerged as a key player in the quantum computing landscape, leading efforts to advance both quantum hardware and software capabilities. The development of scalable quantum processors based on superconducting qubits has been central to IBM Quantum's research and development efforts. Over the years, IBM has made significant strides in increasing qubit counts, improving qubit coherence times, and implementing error correction techniques necessary for reliable quantum computation~\citep{IBM-Berkeley,Utility_3}. These advancements have positioned IBM Quantum as a leader in quantum computing research, with broad implications ranging from computational chemistry and optimization to cryptography and machine learning~\citep{IBM-Berkeley,Utility_3,QSimuIBM1,QSimuIBM,QSimuIBM4,ExxonMobil}.

This paper explores IBM's journey in quantum computing, highlighting key technological achievements, current challenges, and future prospects. The study aims to present a comprehensive review of detailed performance metrics across IBM Quantum's quantum computers, crucial for historical documentation within the NISQ era literature. Metrics examined include 
relaxation times, qubit frequency and anharmonicity, readout assignment errors and readout length, single-qubit gate errors, connection errors, and gate times.

This paper is structured as follows: Section~\ref{S:IBMQ} provides an overview of IBM's quantum computing initiative. 
Section~\ref{S:Processors} examines the progression of IBM Quantum's hardware, detailing advancements from the 5-qubit \textit{Canary} processor to breaking the 1,000-qubit barrier with 
\textit{Condor} processor. Section~\ref{S:Performance} summarizes the performance and characteristics of IBM's quantum computers, covering current systems, their capabilities, as well as retired systems and simulators. 
Moving to Section~\ref{S:Software}, the focus shifts to IBM Quantum's software and tools. Section~\ref{S:Path} outlines IBM Quantum's roadmap and initiatives towards practical quantum computing, including their efforts in quantum safety. 
Finally, Section~\ref{S:Conclusion} provides the conclusion.

%% file: 02IBM.tex
\section{IBM Quantum}\label{S:IBMQ}

\subsection{Overview}

IBM Quantum is a leading provider of quantum computing resources, offering access to top-of-the-line quantum hardware that is built with the latest technology to meet the needs of researchers, industry professionals, and developers.

IBM Quantum operates the most sophisticated collection of quantum systems globally, currently featuring seven utility-scale systems, with additional systems in development~\citep{roadmap-utility}. These systems are known for their exceptional reliability, boasting over 95\% uptime collectively. They also demonstrate remarkable stability, with minimal fluctuations in two-qubit gate errors, which do not exceed 0.001 over periods spanning several months (median 2-qubit gate errors measured across all accessible \textit{Eagle} processors from July 20 to September 20, 2023)~\citep{Eagle’sperformance,Gambetta_utility}

In 2023, IBM has introduced its latest quantum computing milestone with the unveiling of \textit{Condor}, a quantum processor featuring 1,121 superconducting qubits arranged in a honeycomb configuration~\citep{1000-qubit}. This follows the pattern set by earlier record-breaking machines such as \textit{Eagle}, a 127-qubit chip launched in 2021~\citep{100-qubit,Eagle’sperformance,100qubits}, and \textit{Osprey}, a 433-qubit processor announced November 2022 \citep{Ospprey,ibm433}. As part of its strategy, IBM also introduced a new quantum chip named \textit{Heron}, boasting 133 qubits and achieving a remarkable record-low error rate that is three times lower than that of IBM’s previous quantum processor. This achievement marks a departure from IBM's previous strategy of doubling qubit counts annually, signaling a shift towards prioritizing enhanced error resistance over further qubit scalability~\citep{Utility_3,1000-qubit}.

IBM Quantum offers a wide range of hardware and software resources that support learning, experimentation, and collaboration in quantum computing~\citep{IBMQ,Qiskit}, with promising implications for accelerating scientific discovery, enhancing computational efficiency, and addressing complex real-world problems.

\subsection{Breaking the 1,000-qubit barrier}

\begin{figure}
    \centering
    \includegraphics[width=0.41\textwidth]{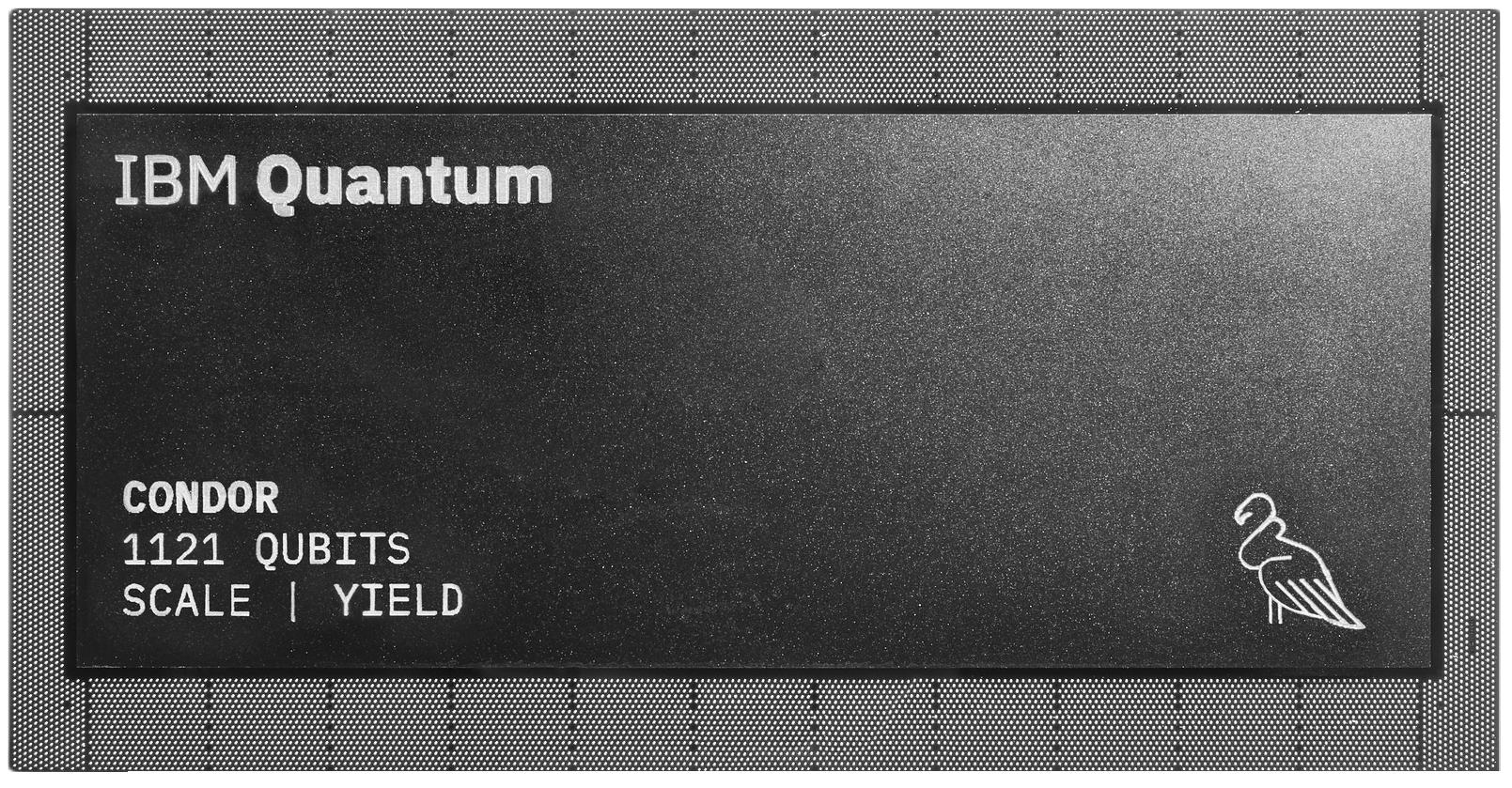}
    \caption{IBM's latest quantum processor \textit{Condor}, unveiled in 2023, features 1,121 superconducting qubits arranged in a honeycomb configuration. (Credit: Ryan Lavine, IBM).}
    \label{fig:condor}
\end{figure}

IBM Quantum has introduced IBM \textit{Condor}, a quantum processor with 1,121 superconducting qubits based on IBM Quantum's cross-resonance gate technology~\citep{1000-qubit}. \textit{Condor} sets new standards in chip design (see Figure~\ref{fig:condor}), featuring a 50\% increase in qubit density, enhancements in qubit fabrication and laminate size, and over a mile of high-density cryogenic flex I/O wiring within a single dilution refrigerator.

Moving to high performing quantum processors,  
IBM Quantum introduced the first IBM Quantum \textit{Heron} processor on the \textit{ibm\_torino} quantum system. Featuring 133 fixed-frequency qubits with tunable couplers. \textit{Heron} delivers significant improvements in device performance (a 3-5x improvement in device performance) compared to IBM Quantum's previous flagship 127-qubit \textit{Eagle} processors~\citep{100-qubit,100qubits,Eagle’sperformance}, while virtually eliminating cross-talk. With \textit{Heron}, IBM Quantum has developed qubit and gate technology that forms the foundation of IBM Quantum’s hardware roadmap moving forward~\citep{Gambetta_utility}.

\subsection{IBM Quantum's system two}

\begin{figure}
    \centering
    \includegraphics[width=0.41\textwidth]{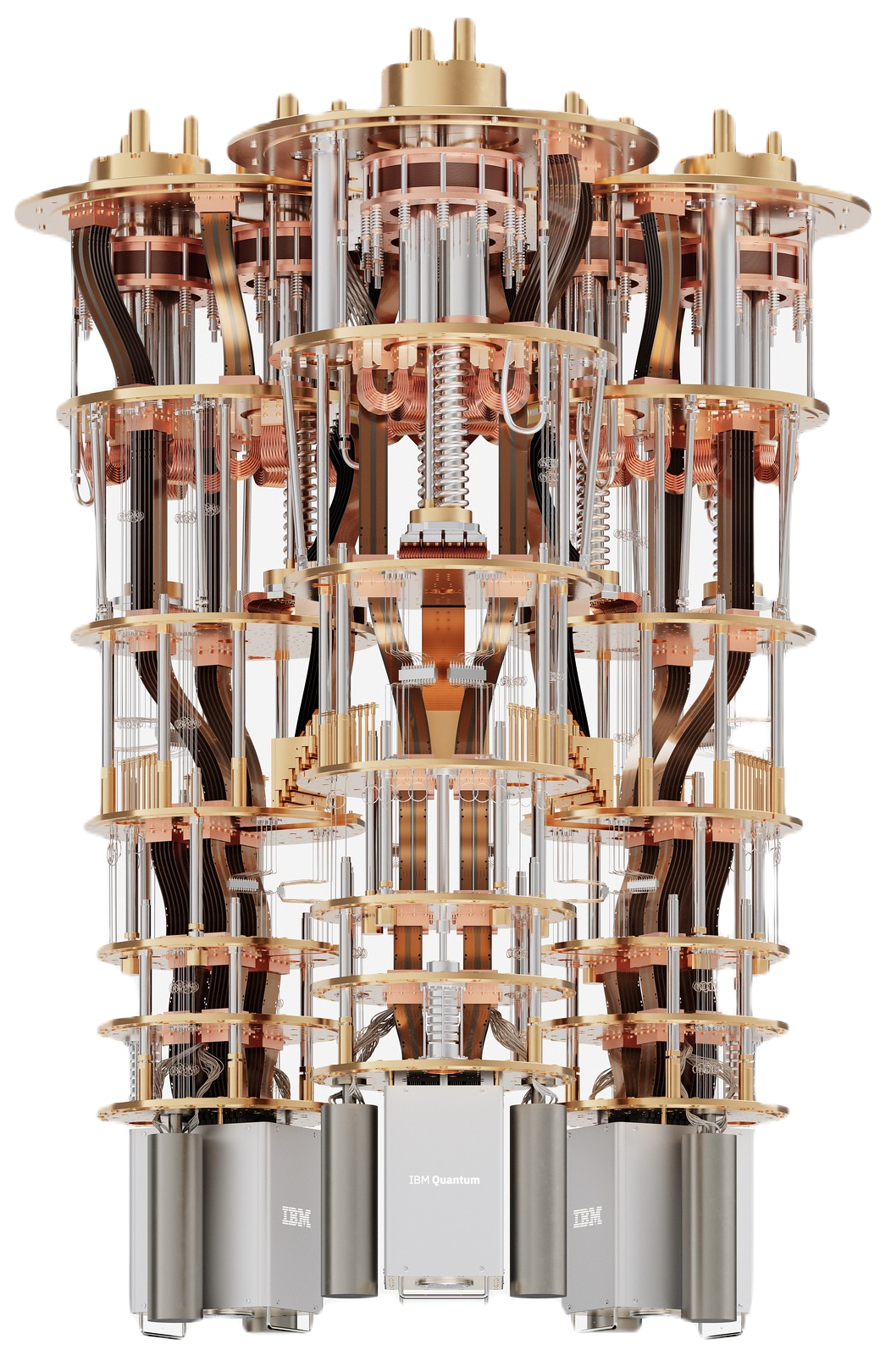}
    \caption{IBM Quantum System Two, unveiled at the IBM Quantum Summit 2023, represents IBM’s first modular quantum computer and serves as a foundational element in IBM’s quantum-centric supercomputing architecture. (Image source: IBM Quantum).}
    \label{fig:SystemTwo}
\end{figure}

IBM Quantum System Two serves as the foundation for scalable quantum computation and is currently operational at the IBM lab in Yorktown Heights, NY. Housing three IBM Quantum \textit{Heron} processors, integrating cryogenic infrastructure with third-generation control electronics and classical runtime servers (see Figure~\ref{fig:SystemTwo}). IBM Quantum System Two features a modular architecture designed to facilitate parallel circuit executions, which are essential for achieving quantum-centric supercomputing~\citep{Gambetta_utility}.

%% file: 03Hardware.tex
\section{The Progression of IBM Quantum's Hardware} \label{S:Processors}

\begin{table*}
    \centering
\caption{Summary of the IBM Quantum’s quantum processors and their progressions.}
\label{Tab:processors}
\vspace{-0.4cm}
    \begin{tabular}{c}
    \includegraphics[width=0.95\textwidth]{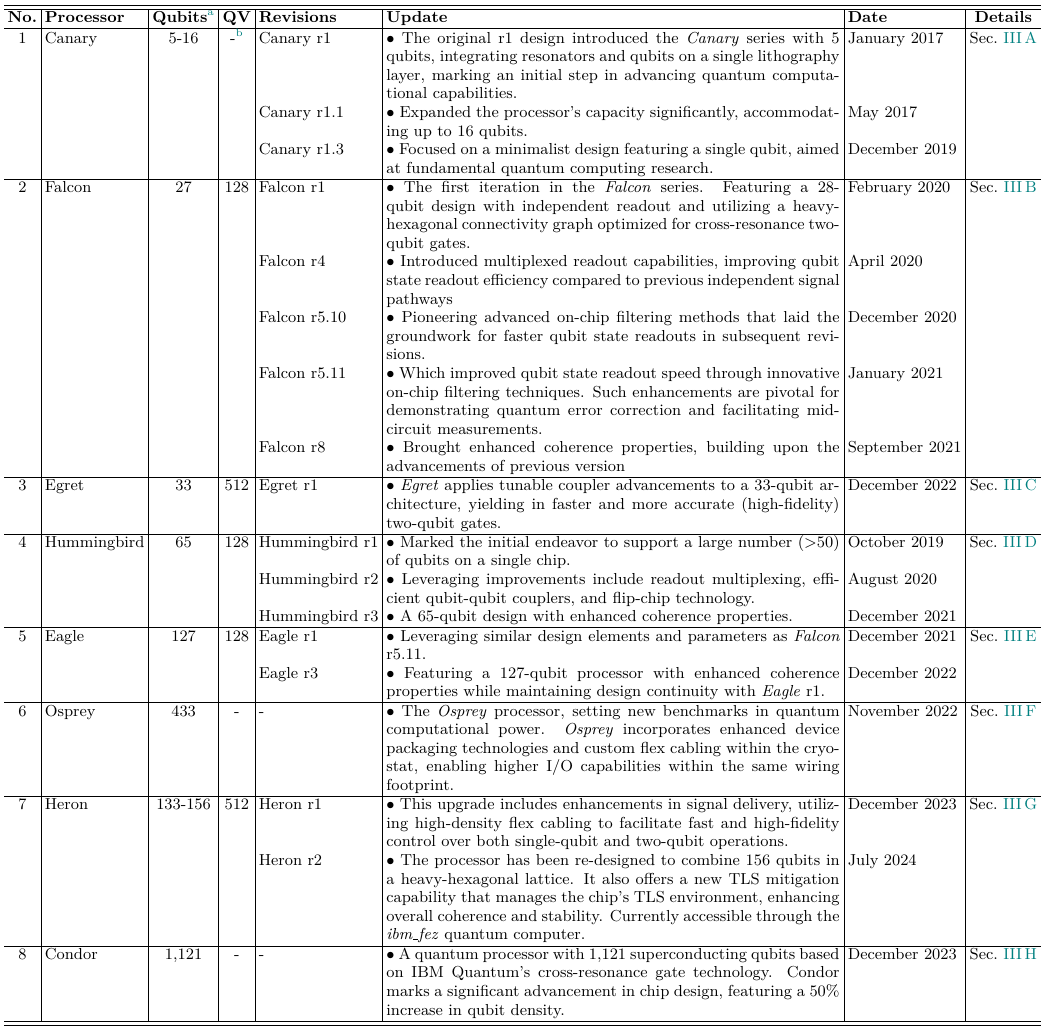} \\
    \end{tabular}
\footnotetext[1]{Maximum number.}
\footnotetext[2]{Value absent from the source.}
\end{table*}

\begin{figure*}
    \centering
    \includegraphics[width=\textwidth]{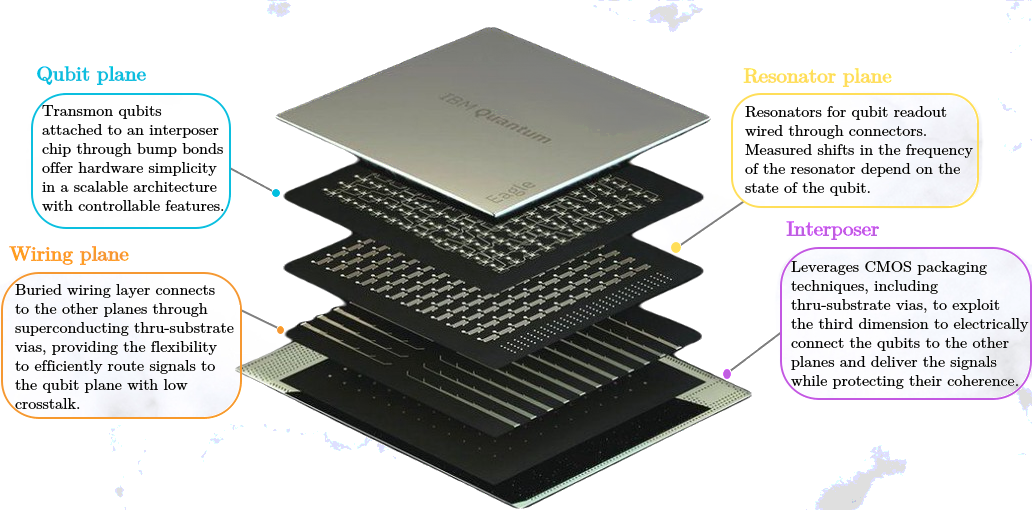}
    \caption{IBM Quantum's \textit{Eagle} processors family configuration. These processors utilize a heavy-hexagonal qubit layout, where qubits connect with two or three neighboring qubits resembling edges and corners of tessellated hexagons. This design reduces errors caused by interactions between adjacent qubits, thereby improving processor reliability and functionality while maintaining high performance.}
    \label{fig:Eagle}
\end{figure*}

IBM Quantum has developed several generations of quantum computers, each contributing to the advancement of quantum computing capabilities. These systems 
are built with cutting-edge superconducting quantum processors based on transmon qubits, chosen for their ability to offer control and scalability~\citep{Stef14}.

Processor types are categorized based on their technological attributes, identified by a combination of family and revision~\citep{Processortypes}. The term ``family" (e.g., \textit{Falcon}) denotes the potential circuit size and complexity achievable on the chip, primarily dictated by the number of qubits and their connectivity structure. ``Revisions" (e.g., r1) signify different design variants within a specific family. These systems represent ongoing advancements in quantum computing hardware, supporting research and development in quantum algorithms and applications. 
In this section, we summarize the progression of IBM Quantum's quantum processors.

\subsection{Canary}\label{S:Canary}

The \textit{Canary} family encompasses compact designs featuring between 5 and 16 qubits, utilizing an optimized 2D lattice where all qubits and readout resonators reside on a single layer~\cite{Processortypes}. 
This family  has seen several revisions aimed at refining and expanding its capabilities. The original r1 design (January 2017) introduced the \textit{Canary} series with 5 qubits, integrating resonators and qubits on a single lithography layer, marking an initial step in advancing quantum computational capabilities. 
Building upon this, r1.1 (May 2017) expanded the processor's capacity, accommodating up to 16 qubits. 
r1.3 (December 2019), which focused on a minimalist design featuring a single qubit, aimed at fundamental quantum computing research~\cite{Processortypes}.

\subsection{Falcon}\label{S:Falcon}

The \textit{Falcon} family is tailored for medium-scale circuits, boasting a QV (quantum volume) of 128. It serves as a crucial testing ground for demonstrating  performance enhancements and scalability improvements before integrating them into larger quantum devices. Native gates and operations supported by \textit{Falcon} devices include \texttt{CX, ID, DELAY, MEASURE, RESET, RZ, SX, X, IF\_ELSE, FOR\_LOOP, \text{and} SWITCH\_CASE.}

Within the \textit{Falcon} family, several revisions have been developed to refine and expand its capabilities: 
The series commenced with \textit{Falcon} r1 in February 2020, characterized by its 28-qubit design with independent readout and utilizing a heavy-hexagonal connectivity graph optimized for cross-resonance two-qubit gates. 
April 2020 saw the introduction of \textit{Falcon} r4, which introduced multiplexed readout capabilities, improving qubit state readout efficiency compared to previous independent signal pathways. \textit{Falcon} r5.10, released in December 2020, pioneering advanced on-chip filtering methods that laid the groundwork for faster qubit state readouts in subsequent revisions. This version also implemented space-saving ``direct-couplers" to enhance qubit coupling efficiency, crucial for scaling quantum systems. \textit{Falcon} r5.11, launched in January 2021, further improved qubit state readout speed through innovative on-chip filtering techniques, facilitating quantum error correction and mid-circuit measurements. In September 2021, the introduction of \textit{Falcon} r8 brought enhanced coherence properties, building upon the advancements of previous versions~\cite{Processortypes}.

\subsection{Egret}\label{S:Egret}

The \textit{Egret} quantum processor features a QV of 512 and introduces tunable couplers on a 33-qubit platform, significantly enhancing the speed and fidelity of two-qubit gates. In December 2022, IBM Quantum launched the first iteration of the \textit{Egret} processor, designated as r1. It demonstrated the highest QV among IBM Quantum systems, marking substantial advancements in reducing two-qubit gate error rates. The \textit{Egret} quantum processor delivers notable improvements in gate fidelity, with many gates achieving 99.9\% fidelity, while also minimizing spectator errors~\cite{Processortypes}.

\subsection{Hummingbird}\label{S:Hummingbird}

The \textit{Hummingbird} family features a QV of 128 and utilizes a heavy-hexagonal qubit layout, accommodating up to 65 qubits. October 2019 marked the debut of \textit{Hummingbird} r1, representing the initial effort to support a large number ($>$50) of qubits on a single chip, setting the stage for subsequent advancements in quantum processor design and scalability. August 2020 saw the release of \textit{Hummingbird} r2, featuring 65 qubits and leveraging improvements, 
such as readout multiplexing, efficient qubit-qubit couplers, and flip-chip technology, which collectively enhance the scalability and operational capabilities of the \textit{Hummingbird} family. In December 2021, \textit{Hummingbird} r3 introduced a 65-qubit design with enhanced coherence properties, reflecting advancements in quantum processing stability and performance~\cite{Processortypes}.

\subsection{Eagle}\label{S:Eagle}

\begin{figure*}[htb]
    \centering
    \includegraphics[width=\textwidth]{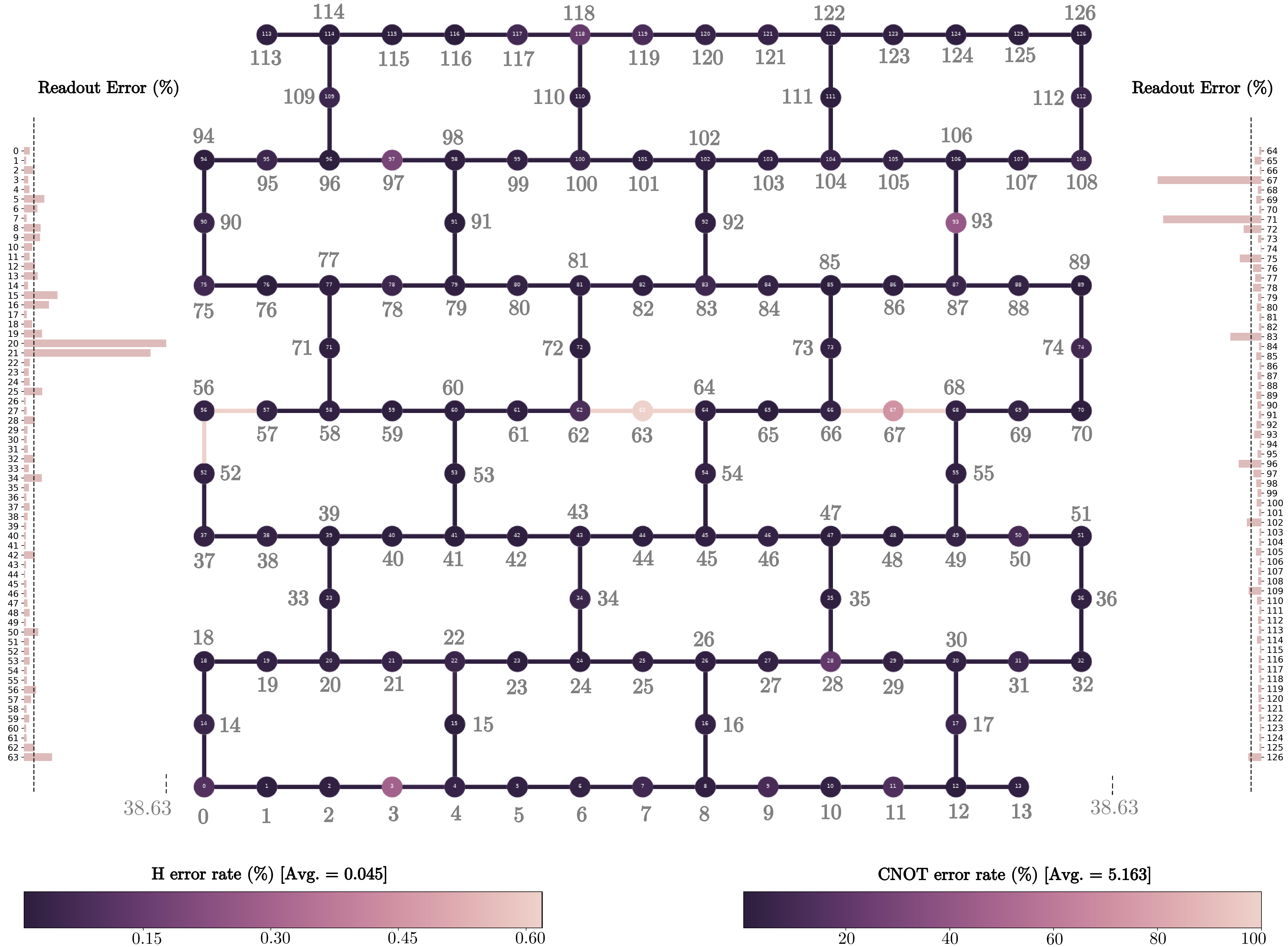}
\caption{Readout error map and layout of the \textit{ibm\_sherbrooke} quantum computer. This quantum system is based on an \textit{Eagle r3} quantum processor, featuring 127 superconducting transmon qubits. Key performance metrics include 
median ECR error: $7.571 \times 10^{-3}$, 
median SX error: $2.411  \times 10^{-4}$
median readout error: $1.350  \times 10^{-2}$
median T1: 262.69 $\mu$s, and 
median T2: 176.67 $\mu$s,  
as of August 1, 2024. }
\label{fig:sherbrooke}
\end{figure*}

The \textit{Eagle} family boasts a QV of 128 and integrates advanced packaging technologies to accommodate 127 qubits~\citep{100-qubit}. These processors employ a heavy-hexagonal qubit layout, where qubits are connected to two or three neighbors, resembling the edges and corners of tessellated hexagons~\citep{100qubits}, see Figure~\ref{fig:Eagle}. This layout minimizes errors from interactions between adjacent qubits, thereby enhancing processor reliability and functionality without compromising performance~\citep{Eagle’sperformance}. Native gates and operations supported include 
\texttt{ECR, ID, DELAY, MEASURE, RESET, RZ, SX, X, IF\_ELSE, FOR\_LOOP, AND SWITCH\_CASE.} 

In December 2021, \textit{Eagle} r1 was introduced, leveraging similar design elements and parameters as \textit{Falcon} r5.11. This version supports fast qubit readout and aims for comparable gate speeds and error rates~\citep{Eagle’sperformance}. 
In December 2022, \textit{Eagle} r3 was released, featuring a 127-qubit processor with enhanced coherence properties while maintaining design continuity with \textit{Eagle} r1~\cite{Processortypes}. Currently, \textit{Eagle} systems are deployed at various IBM Quantum's quantum computers. The error map and layout of a quantum computer that is built based on an \textit{Eagle} processor, such as \textit{ibm\_sherbrooke}, is depicted in Figure~\ref{fig:sherbrooke}.

\subsection{Osprey}\label{S:Osprey}

The \textit{Osprey} quantum processor boasting 433 qubits, nearly four times the size of its predecessor, \textit{Eagle}. 
\textit{Osprey} integrates enhanced device packaging technologies and custom flex cabling within the cryostat, enabling higher I/O capabilities within the same wiring footprint. IBM Quantum unveiled the \textit{Osprey} processor in November 2022. \textit{Osprey} has the potential to execute complex computations far surpassing the capabilities of classical computers. For context, the number of classical bits required to represent a single state on the IBM \textit{Osprey} processor exceeds the total number of atoms in the observable universe~\cite{Processortypes}.

\subsection{Heron}\label{S:Heron}

\textit{Heron} represents a significant advancement in quantum computing, featuring a QV of 512 and incorporating innovations in signal delivery previously seen in the \textit{Osprey} processor. With 133 qubits, \textit{Heron} builds upon the size and capabilities of its predecessor, \textit{Egret}, and shares a similar footprint to \textit{Eagle}. This upgrade includes enhancements in signal delivery, utilizing high-density flex cabling to facilitate fast and high-fidelity control over both single-qubit and two-qubit operations. Native gates and operations supported by \textit{Heron} include \texttt{CZ, ID, DELAY, MEASURE, RESET, RZ, SX, X, IF\_ELSE, FOR\_LOOP, and SWITCH\_CASE.}

In December 2023, \textit{Heron} r1 was introduced, leveraging 133 qubits as the first version of \textit{Heron}, currently accessible through \textit{ibm\_torino}. In July 2024, the processor has been re-designed to combine 156 qubits in a heavy-hexagonal lattice. It also offers a new TLS mitigation capability that manages the chip’s TLS environment, enhancing overall coherence and stability. The \textit{Heron} r2 processor, currently accessible through
\textit{ibm\_fez} quantum computer~\cite{Processortypes}.

\subsection{Condor}\label{S:Condor}

In December 2023, IBM Quantum unveiled \textit{Condor}, a groundbreaking quantum processor consists of 1,121 superconducting qubits and leveraging IBM Quantum's
cross-resonance gate technology~\citep{1000-qubit,cross-resonance}. This technology facilitates precise two-qubit operations between superconducting qubits with fixed frequencies, known for their simplicity in implementation and resilience against noise~\citep{cross-resonance}.

\textit{Condor} represents a leap forward in chip design, boasting a 50\% increase in qubit density and notable enhancements in qubit fabrication and laminate size. Moreover, it integrates an impressive length of over a mile of high-density cryogenic flex I/O wiring within a single dilution refrigerator. With performance comparable to its predecessor, the 433-qubit \textit{Osprey}, \textit{Condor} signifies a significant milestone in quantum computing innovation. It effectively tackles scalability challenges while offering valuable insights for future hardware designs~\cite{Processortypes}.

%% file: 04Performance.tex
\section{Performance and Characteristics of IBM's Quantum Computers} \label{S:Performance}

\subsection{Evolution of Quantum Systems}

IBM Quantum's quantum computing offerings have evolved to include both current and retired systems. 
These systems range from early developments to advanced processors with up to 433 qubits (see  Table~\ref{tab:retierd_systems} for specific details). The journey from retired systems to the latest generations reflects IBM's continuous efforts to push the boundaries of quantum computation (see Table~\ref{ibmsystems}).  
For a view of the availability and details of IBM Quantum's current quantum systems, including access plans, interested readers are referred to~\citep{QPU}.

\subsection{Retired systems and simulators}

IBM's commitment to quantum computing is evident through a series of pioneering systems, some of which have since been retired. Systems such as \textit{ibmq\_5\_yorktown} and \textit{ibmq\_16\_melbourne}, retired on August 9, 2021. 
\textit{ibmq\_manhattan} followed on September 22, 2021, and earlier systems such as \textit{ibmq\_athens} and \textit{ibmq\_rome} were retired on June 30, 2021. Refer to Table~\ref{tab:retierd_systems} for a list of retired systems. 
These quantum systems, alongside quantum simulators, were pivotal in early quantum algorithm experimentation and have paved the way for newer generations that continue to push the boundaries of quantum computing. Appendix~\ref{appendix:A1} provides detailed performance summaries of these quantum computers, recorded for historical documentation in the NISQ computing era literature.

\begin{table*}
    \centering
\caption{List of IBM Quantum's retired quantum systems and cloud simulators. Older systems are identified with names starting with “ibmq,” while newer systems use names beginning with “ibm.”} 
\label{tab:retierd_systems} 
    \begin{tabular}{c}
    \includegraphics[width=\textwidth]{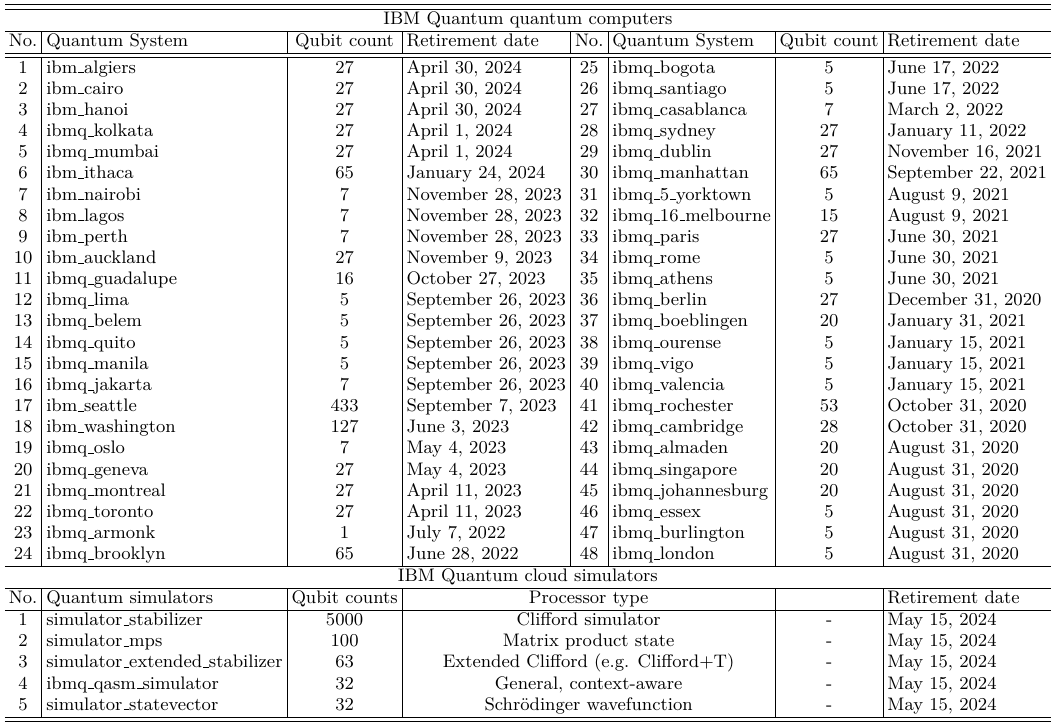} \\
    \end{tabular}
\end{table*}

\begin{table*}
    \centering
\caption{The up-to-date IBM Quantum’s quantum computers.} 
\label{ibmsystems}
    \begin{tabular}{c}
    \includegraphics[width=\textwidth]{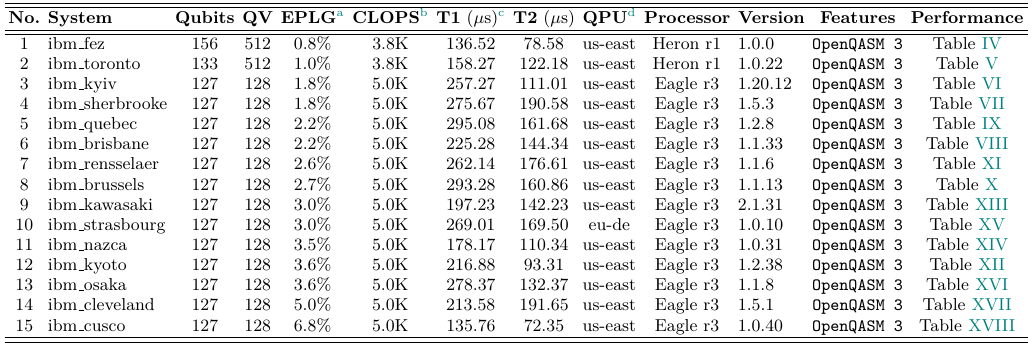} \\
    \end{tabular}
\footnotetext[1]{Error per layered gate for a 100-qubit chain.}
\footnotetext[2]{Hardware-aware circuit layer operations per second.} 
\footnotetext[3]{The median values of the relaxation time-T1 and the coherence time-T2, measured in
microseconds ($\mu$s). Accessed July 2, 2024. except for the \textit{ibm\_fez} accessed 04 July 2024. Most systems support a maximum of 300 circuits and 100,000 shots.}
\footnotetext[4]{QPU region.}
\end{table*}

\subsection{The up-to-date machines' performance}

This section analyzes the performance metrics of 15 up-to-date IBM Quantum's quantum machines. 
Tables~\ref{tt:fez} to~\ref{tt:cusco} summarize the hardware performance, qubit characteristics, and specifications of these quantum computers. Key parameters such as coherence times (T1 and T2), qubit frequencies, qubit anharmonicity, readout assignment error, readout length, qubit flip probabilities, as well as error rates for both single-qubit and two-qubit gates.

\begin{table*}
    \centering
\caption{Summary of hardware performance, qubit characteristics, and key specifications for the 156 qubit quantum computer \textit{ibm\_fez}. 
The basis gates employed by this system are CZ, ID, RZ, SX, and X.  
The processor type is \textit{Heron} r2 (version 1.0.0). 
As of July 4, 2024, the system demonstrates a median CZ error of $2.848 \times 10^{-3}$, 
and a median gate time of 68 ns.}
\label{tt:fez} 
    \begin{tabular}{c}
    \includegraphics[width=\textwidth]{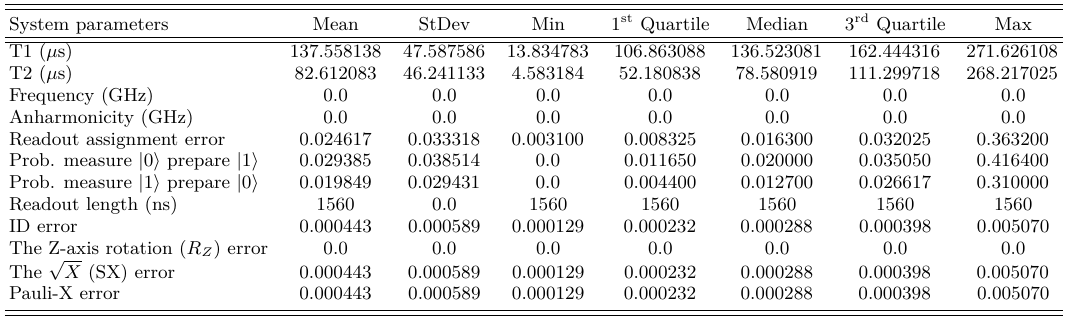} \\
    \end{tabular}
\end{table*}

\begin{table*}
    \centering
\caption{Summary of hardware performance, qubit characteristics, and key specifications for the 133 qubit quantum computer \textit{ibm\_torino}. 
The basis gates of this machine are CZ, ID, RZ, SX, and X.  
The processor type is \textit{Heron} r1 (version  1.0.22). 
As of July 3, 2024, the system demonstrates a median CZ error of $4.769 \times 10^{-3}$, 
and a median gate time of 84 ns.}
\label{tt:torino}  
    \begin{tabular}{c}
    \includegraphics[width=\textwidth]{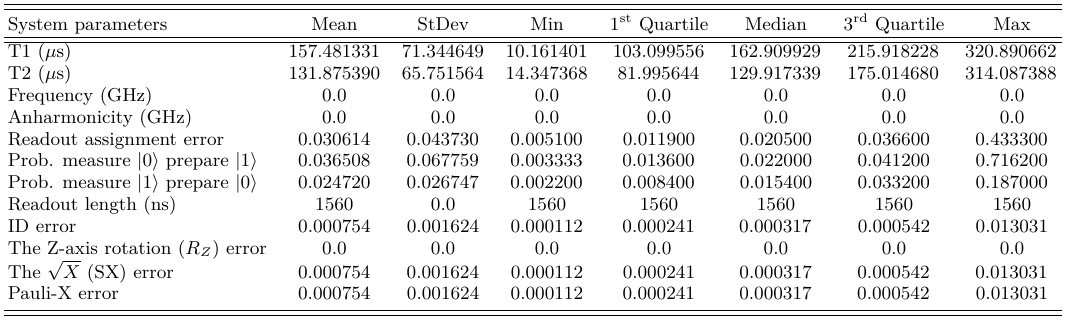} \\
    \end{tabular}
\end{table*}

\begin{table*}
    \centering
\caption{Summary of hardware performance, qubit characteristics, and key specifications for the 127-qubit quantum computer \textit{ibm\_kyiv}. 
The basis gates employed by this system include ECR, RZ, SX, ID, and X.   
The processor type is \textit{Eagle} r3 (version 1.20.12). 
As of July 3, 2024, the system demonstrates a median ECR error of $1.160 \times 10^{-2}$, 
and a median gate time of 561.778 ns.}
\label{tt:kyiv} 
    \begin{tabular}{c}
    \includegraphics[width=\textwidth]{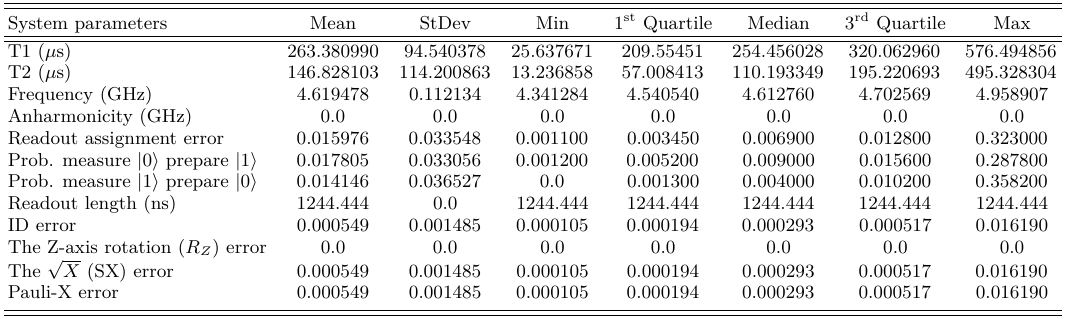} \\
    \end{tabular}
\end{table*}

\begin{table*}
    \centering
\caption{Summary of hardware performance, qubit characteristics, and key specifications for the 127-qubit quantum computer \textit{ibm\_sherbrooke}. 
The basis gates employed by this system include ECR, RZ, SX, ID, and X.    
The processor type is  Eagle r3 (version 1.5.3). 
As of July 3, 2024, the system demonstrates a median ECR error of $7.400 \times 10^{-3}$, 
and a median gate time of 533.333 ns.}
\label{tt:sherbrooke} 
    \begin{tabular}{c}
    \includegraphics[width=\textwidth]{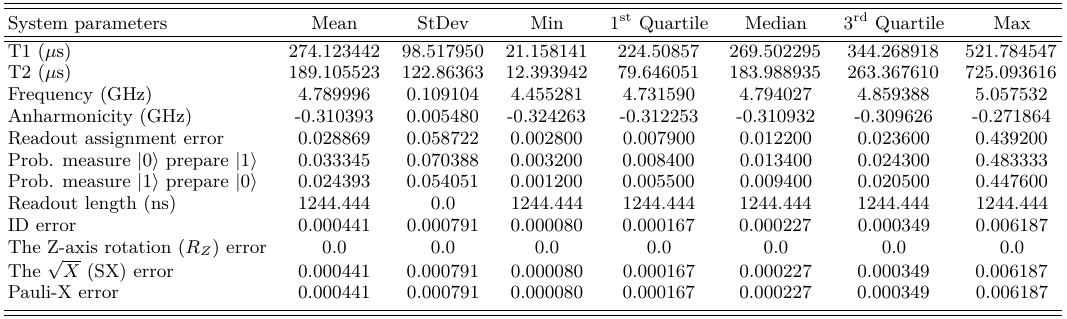} \\
    \end{tabular}
\end{table*}

\begin{table*}
    \centering
\caption{Summary of hardware performance, qubit characteristics, and key specifications for the 127-qubit quantum computer \textit{ibm\_brisbane}. 
The basis gates employed by this system include ECR, RZ, SX, ID, and X. 
The processor type is \textit{Eagle} r3 (version 1.1.33). 
As of July 3, 2024, the system demonstrates a median ECR error of $8.335 \times 10^{-3}$, 
and a median gate time of 660 ns.}
\label{tt:brisbane} 
    \begin{tabular}{c}
    \includegraphics[width=\textwidth]{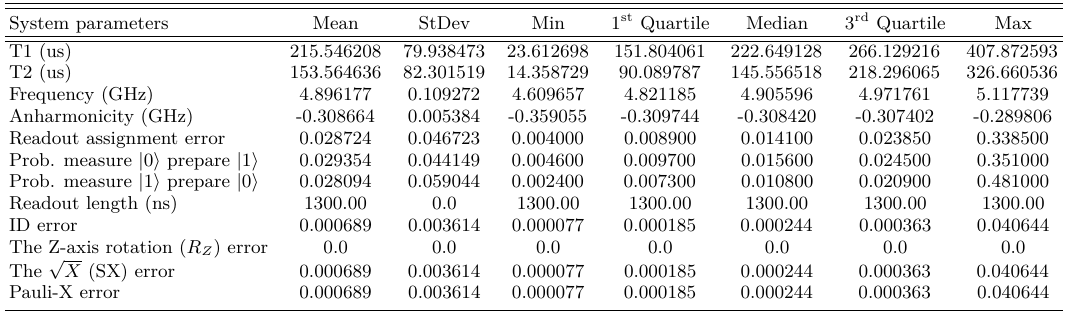} \\
    \end{tabular}
\end{table*}

\begin{table*}
    \centering
\caption{Summary of hardware performance, qubit characteristics, and key specifications for the 127-qubit quantum computer \textit{ibm\_quebec}. 
The basis gates of this machine are  ECR, RZ, SX, ID, X. 
The processor type is \textit{Eagle} r3 (version 1.2.8). 
As of July 3, 2024, the system demonstrates a median ECR error of $8.017 \times 10^{-3}$, 
and a median gate time of 593 ns.}
\label{tt:quebec} 
    \begin{tabular}{c}
    \includegraphics[width=\textwidth]{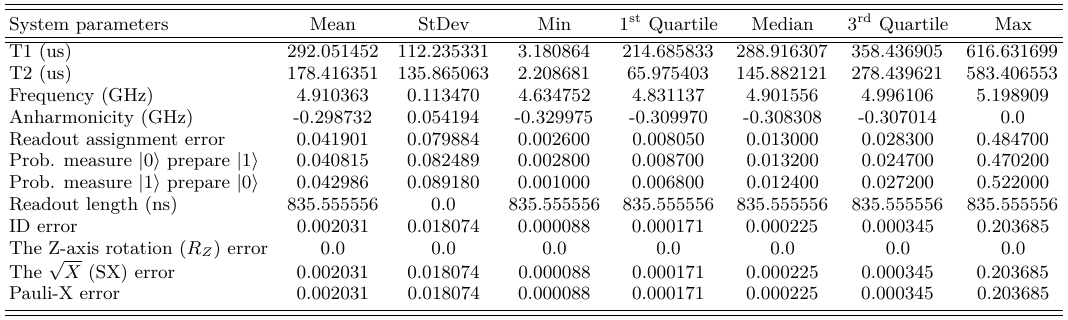} \\
    \end{tabular}
\end{table*}

\begin{table*}
    \centering
\caption{Summary of hardware performance, qubit characteristics, and key specifications for the 127-qubit quantum computer \textit{ibm\_brussels}. 
The basis gates employed by this system include ECR, RZ, SX, ID, and X.   
The processor type is \textit{Eagle} r3 (version 1.1.13). 
As of July 3, 2024, the system demonstrates a median ECR error of $8.074 \times 10^{-3}$, 
and a median gate time of 660 ns.}
\label{tt:brussels} 
    \begin{tabular}{c}
    \includegraphics[width=\textwidth]{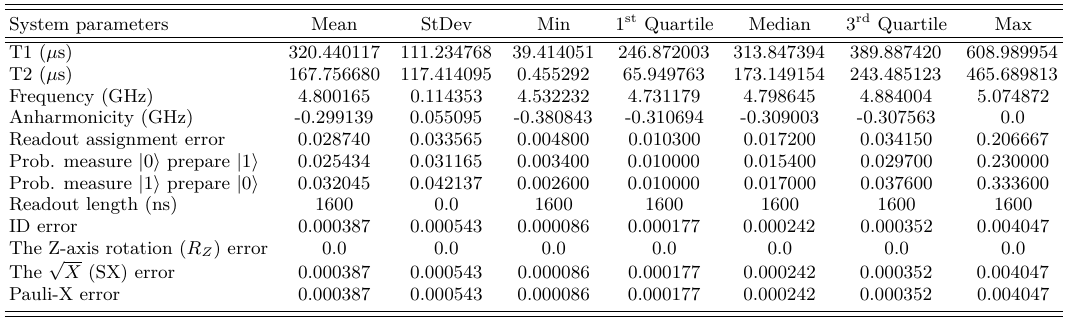} \\
    \end{tabular}
\end{table*}

\begin{table*}
    \centering
\caption{Summary of hardware performance, qubit characteristics, and key specifications for the 127-qubit quantum computer, \textit{ibm\_rensselaer}. 
The basis gates employed by this system include ECR, RZ, SX, ID, and X.   
The processor type is \textit{Eagle} r3 (version 1.1.6). 
As of July 3, 2024, this quantum system exhibits a median  ECR error: $7.580 \times 10^{-3}$, 
and a median gate time of 665.889 ns.}
\label{tt:rensselaer} 
    \begin{tabular}{c}
    \includegraphics[width=\textwidth]{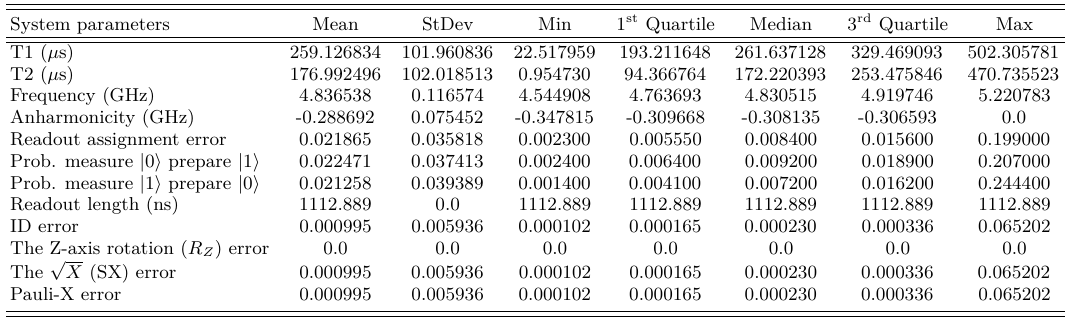} \\
    \end{tabular}
\end{table*}

\begin{table*}
    \centering
\caption{Summary of hardware performance, qubit characteristics, and key specifications for the 127-qubit quantum computer \textit{ibm\_kyoto}. 
The basis gates employed by this system include ECR, RZ, SX, ID, and X.   
The processor type is \textit{Eagle} r3 (version 1.2.38). 
As of July 3, 2024, the system demonstrates a median ECR error of $1.023 \times 10^{-2}$, 
and a median gate time of 660 ns.}
\label{tt:kyoto} 
    \begin{tabular}{c}
    \includegraphics[width=\textwidth]{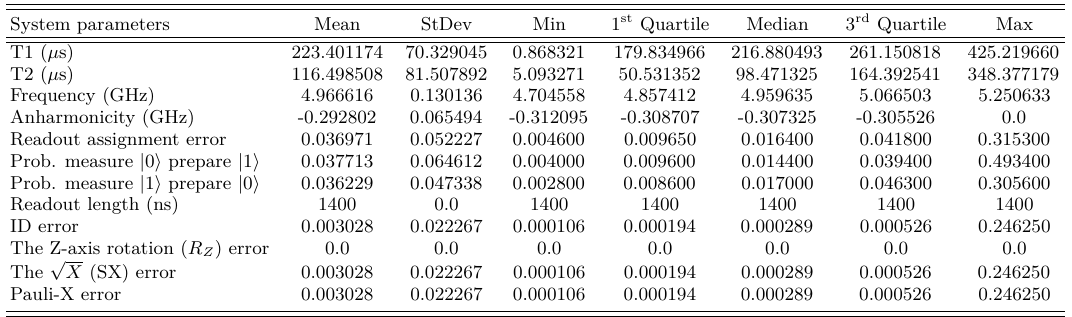} \\
    \end{tabular}
\end{table*}

\begin{table*}
    \centering
\caption{Summary of hardware performance, qubit characteristics, and key specifications for the 127-qubit quantum computer \textit{ibm\_kawasaki}. 
The basis gates employed by this system include ECR, RZ, SX, ID, and X.   
The processor type is \textit{Eagle} r3 (version 2.1.31). 
As of July 3, 2024, the system demonstrates a median ECR error of $7.114 \times 10^{-3}$, 
and a median gate time of 586.667 ns.}
\label{tt:kawasaki} 
    \begin{tabular}{c}
    \includegraphics[width=\textwidth]{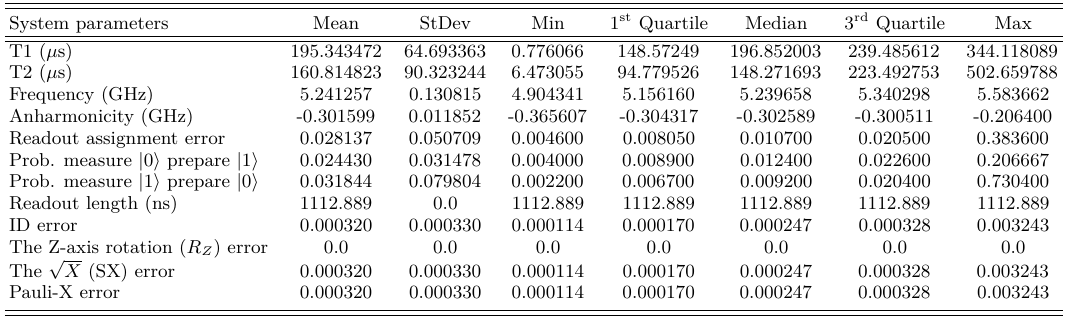} \\
    \end{tabular}
\end{table*}

\begin{table*}
    \centering
\caption{Summary of hardware performance, qubit characteristics, and key specifications for the 127-qubit quantum computer \textit{ibm\_nazca}. 
The basis gates employed by this system include ECR, RZ, SX, ID, and X.   
The processor type is \textit{Eagle} r3 (version 1.0.31). 
As of July 3, 2024, the system demonstrates a median ECR error of $1.240 \times 10^{-2}$, 
and a median gate time of 660 ns.}
\label{tt:nazca} 
    \begin{tabular}{c}
    \includegraphics[width=\textwidth]{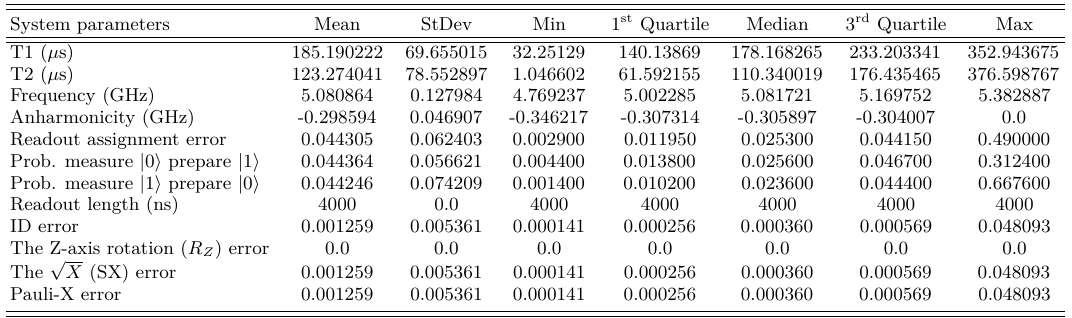} \\
    \end{tabular}
\end{table*}

\begin{table*}
    \centering
\caption{Summary of hardware performance, qubit characteristics, and key specifications for the 127-qubit quantum computer \textit{ibm\_strasbourg}. 
The basis gates of this machine are  ECR, RZ, SX, ID, X.  
The processor type is \textit{Eagle} r3 (version 1.10.11). 
As of July 3, 2024, the system demonstrates a median ECR error of $8.857 \times 10^{-3}$, 
and a median gate time of 660 ns.}
\label{tt:strasbourg} 
    \begin{tabular}{c}
    \includegraphics[width=\textwidth]{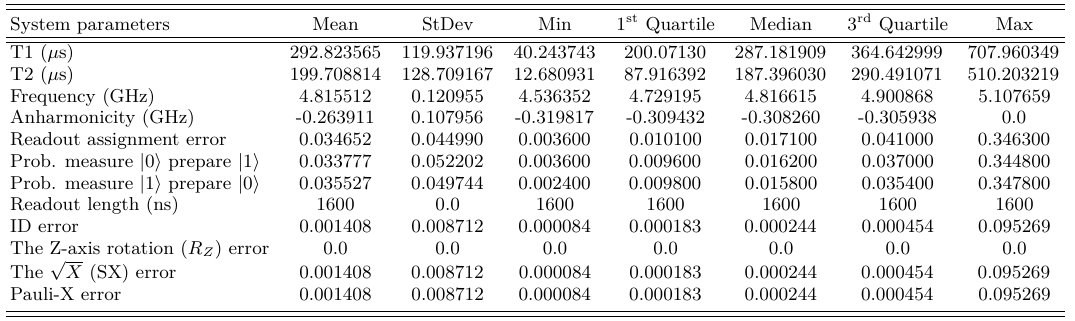} \\
    \end{tabular}
\end{table*}

\begin{table*}
    \centering
\caption{Summary of hardware performance, qubit characteristics, and key specifications for the 127-qubit quantum computer \textit{ibm\_osaka}. 
The basis gates employed by this system include ECR, RZ, SX, ID, and X.   
The processor type is \textit{Eagle} r3 (version 1.1.8). 
As of July 3, 2024, the system demonstrates a median ECR error of $8.596 \times 10^{-3}$, 
and a median gate time of 660 ns.}
\label{tt:osaka} 
    \begin{tabular}{c}
    \includegraphics[width=\textwidth]{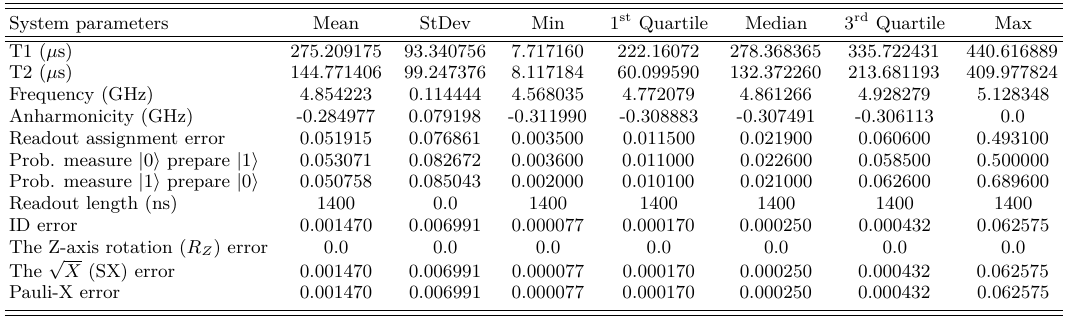} \\
    \end{tabular}
\end{table*}

\begin{table*}
    \centering
\caption{Summary of hardware performance, qubit characteristics, and key specifications for the 127-qubit quantum computer \textit{ibm\_cleveland}. 
The basis gates of this machine are  ECR, RZ, SX, ID, X.  
The processor type is \textit{Eagle} r3 (version 1.5.1). 
As of July 3, 2024, the system demonstrates a median ECR error of $9.157 \times 10^{-3}$, 
and a median gate time of 590.222 ns.}
\label{tt:cleveland}  
    \begin{tabular}{c}
    \includegraphics[width=\textwidth]{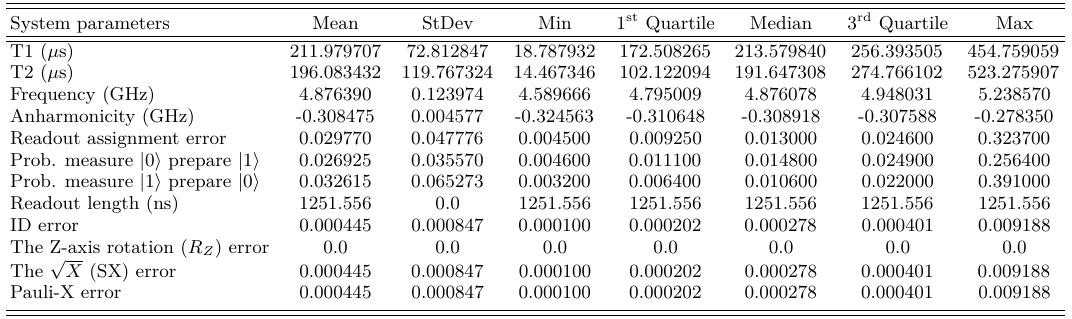} \\
    \end{tabular}
\end{table*}

\begin{table*}
    \centering
\caption{Summary of hardware performance, qubit characteristics, and key specifications for the 127-qubit quantum computer \textit{ibm\_cusco}. 
The basis gates employed by this system include ECR, RZ, SX, ID, and X.   
The processor type is \textit{Eagle} r3 (version 1.0.40). 
As of July 3, 2024, the system demonstrates a median ECR error of $2.398 \times 10^{-2}$, 
and a median gate time of 460 ns.}
\label{tt:cusco} 
    \begin{tabular}{c}
    \includegraphics[width=\textwidth]{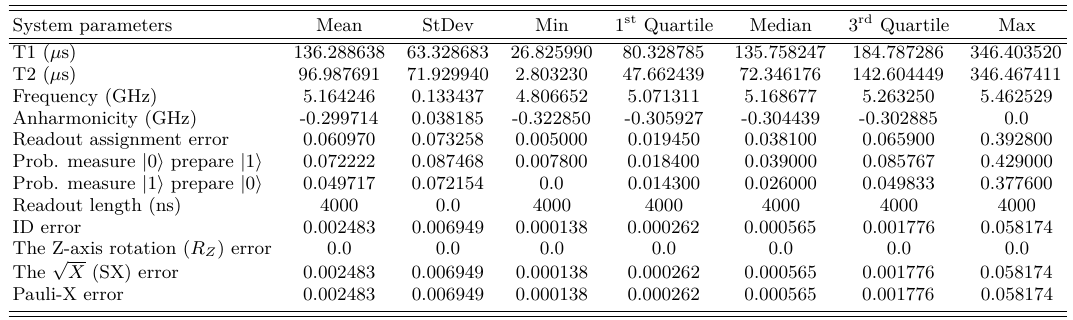} \\
    \end{tabular}
\end{table*}

%% file: 05Software.tex
\section{The IBM Quantum's Software} \label{S:Software} 

\begin{figure*}
    \centering
      \includegraphics[width=\textwidth]{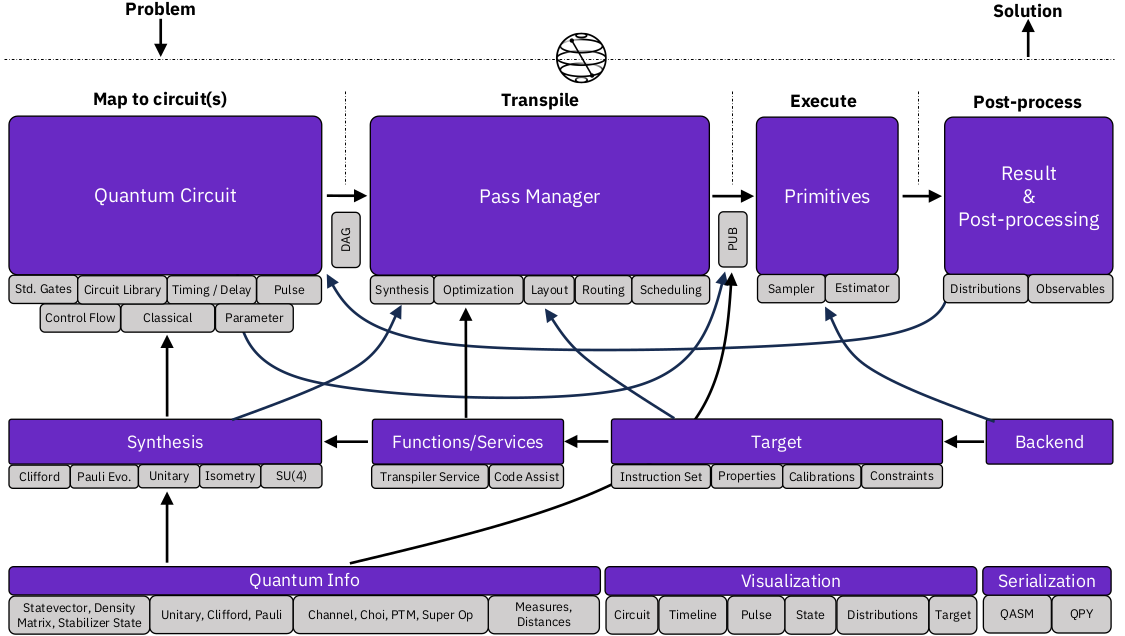}
     \caption{
     Qiskit's software architecture: The quantum info module links circuits to quantum information mathematics. The transpiler optimizes circuits via a pass manager, considering ISA and constraints. Primitives run circuits on simulators or hardware, evaluating results. Visualization tools and serialization with OpenQASM~\citep{QCQiskit31} and QPY format are also included~\citep{QCQiskit54}. Regenerated under a Creative Commons license (\url{ https://creativecommons.org/licenses/by/4.0/}) from~\citep{QCQiskit}.
     }
    \label{fig:Qiskit2}
\end{figure*}

\begin{figure*}
    \centering
    \includegraphics[width=\textwidth]{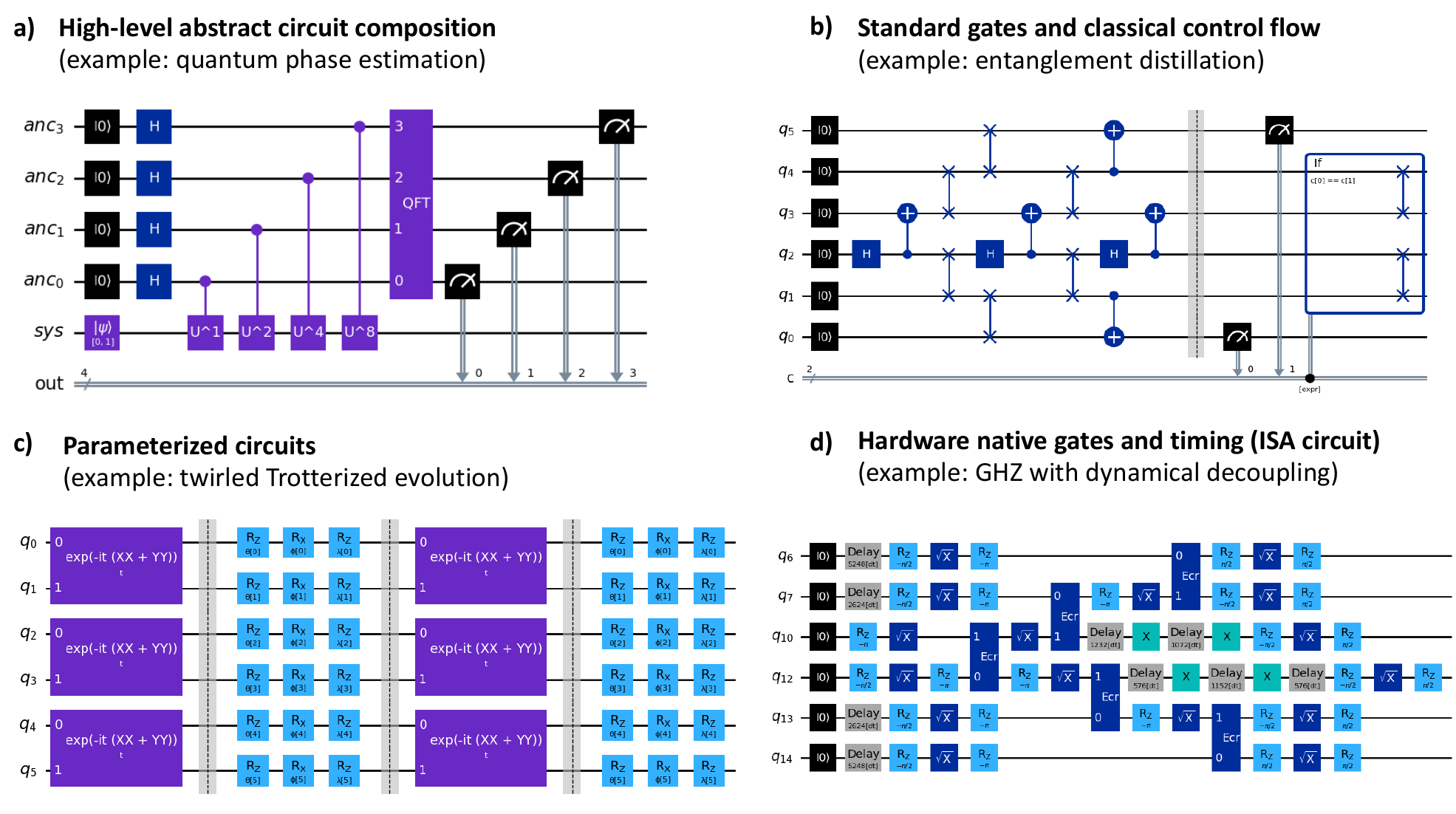}
    \caption{
    Examples of Qiskit circuits: (a) Quantum phase estimation algorithm~\citep{QCQiskit53}. 
    (b) Entanglement distillation circuit with fallback logic for Bell state preparation~\citep{QCQiskit23}. 
    (c) Trotterized $XX + YY$ Hamiltonian simulation  circuit~\citep{QCQiskit58} with Pauli twirling for noise suppression~\citep{QCQiskit22}. 
    (d) GHZ state preparation circuit optimized for hardware ISA, using dynamical decoupling to reduce noise~\citep{QCQiskit95}. Regenerated under a Creative Commons license (\url{ https://creativecommons.org/licenses/by/4.0/}) from~\citep{QCQiskit}.
    } 
    \label{fig:Qiskit3} 
\end{figure*}

\subsection{Qiskit}

Qiskit, the software development kit for quantum information science, launched by IBM in 2017 as an open-source toolbox for quantum computing. Over the past six years, it has flourished significantly. Qiskit has been installed over 6 million times, with current installations occurring at a rate of 300,000 per month~\citep{QCQiskit}. Boasting more than 2,000 forks, over 8,000 contributions, and has facilitated the execution of over 3 trillion circuits~\citep{QCQiskit}. By a significant margin, Qiskit stands out as the most widely-adopted quantum computing software~\cite{QCQiskit10}.

Qiskit has demonstrated its effectiveness in recent studies in quantum computing, particularly in error mitigation~\cite{IBM-Berkeley}. Moreover, it played a crucial role in achieving fault-tolerant magic state preparation surpassing break-even fidelity~\cite{QCQiskit42}, as well as in numerous significant studies involving up to 133 qubits and thousands of two-qubit entangling gates~\cite{Entangling,QCQiskit21,QCQiskit40,SQSCZ1,QCQiskit30,QCQiskit39,QCQiskit59,SQSCZ2,QCQiskit65,QCQiskit72,GSA,QCQiskit101,QCQiskit100,EPJQ,Googles,QCQiskit104}.

\subsection{Qiskit patterns}

Qiskit patterns, outline a structured four-step process for executing algorithms on quantum computers, aligning with its software architecture (shown in Figure~\ref{fig:Qiskit2}). 
Initially, classical problems are translated into quantum computations by constructing circuits that encode the specific problem. Qiskit provides a user-friendly circuit construction API capable of handling extensive circuits. 
Subsequently, circuits undergo transformation–referred to as transpilation–to optimize them for execution on target hardware, focusing on circuit-to-circuit rewriting rather than full compilation to classical controller instructions. 
Following transpilation, circuits are executed on a target backend using primitive computations. Finally, the obtained results are post-processed to derive solutions for the original problem.

Workflows may iterate through these steps, incorporating advanced patterns like 
generating new circuits based on results from prior batches~\citep{QCQiskit29},  
integrating quantum and classical computing in a quantum-centric supercomputing architecture~\citep{QCQiskit16,QCQiskit78}. 
Complex pattern orchestration is streamlined via the Qiskit serverless framework~\citep{QCQiskit38}.

\subsection{Qiskit circuits}

Quantum circuits are central to Qiskit's architecture, representing computations as sequences of instructions that can be manipulated and analyzed within the software. Qiskit defines circuits broadly, encompassing operations on both quantum and classical data. This includes standard actions like qubit operations and measurements, as well as advanced mathematical operators such as unitaries, Cliffords, isometries, and Fourier transforms. Circuits may also involve classical computations in real-time, such as applying Boolean functions to measurement outcomes, and classical control flow mechanisms like loops and branches.

Circuits can also delineate timing operations and continuous-time qubit dynamics (CTQD) using pulse-defined gates. These levels of abstraction can be combined within a single circuit, facilitating modular composition. Figure~\ref{fig:Qiskit2} provides an overview of Qiskit’s architecture, highlighting its components and interactions, while 
Figure~\ref{fig:Qiskit3} illustrates diverse circuit types supported by Qiskit~\citep{QCQiskit}. This flexibility supports the exploration of various quantum algorithms and physical implementations.

\subsection{Scale to large numbers of qubits}

In quantum computing, advancing in the field requires tackling utility-scale tasks, which involve computations on a significantly larger scale. This entails working with circuits that utilize more than 100 qubits and incorporate over 1000 gates.

To demonstrate large-scale operations on IBM Quantum systems, we consider the following example, involving the generation and analysis of a 100-qubit GHZ state (\(  |\text{GHZ}\rangle_{100} = \frac{1}{\sqrt{2}} (|0\rangle^{\otimes 100} + |1\rangle^{\otimes 100}) \))~\citep{scale_qiskit}. This approach leverages the Qiskit patterns workflow and concludes with the measurement of the expectation value \( \langle Z_0 Z_i \rangle \) for each qubit. The process of developing a quantum program using Qiskit entails four essential steps: mapping the problem to a quantum-native format, optimizing circuits and operators, executing with a quantum primitive function, and analyzing the resulting data. 

\newpage

$\bullet$ 
\textit{Mapping the Problem}. Begin by constructing a function that generates a \texttt{QuantumCircuit} 
specifically designed to prepare an n-qubit GHZ state. Then, apply this function to generate a 100-qubit GHZ state and collect the relevant observables for measurement.

\begin{table}[H]
    \centering
    \begin{tabular}{c}
    \includegraphics[width=0.47\textwidth]{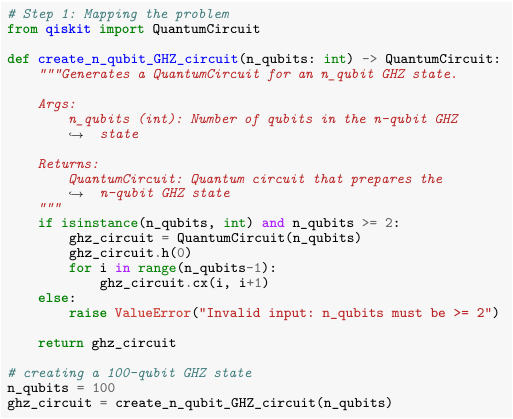} \\
    \end{tabular}
    \captionsetup{labelformat=empty, labelsep=none} 
\end{table}

Next, proceed to map to the operators of interest. In this case, the focus is on $ZZ$ operators between qubits to analyze their behavior over increasing distances. The goal  is to observe how expectation values become progressively less accurate (more corrupted), indicating the extent of noise present in the system.

\begin{table}[H]
    \centering
    \begin{tabular}{c}
    \includegraphics[width=0.47\textwidth]{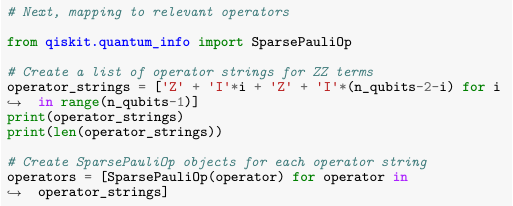} \\
    \end{tabular}
    \captionsetup{labelformat=empty, labelsep=none} 
\end{table}

$\bullet$ \textit{Optimization for quantum hardware execution.} Optimize the problem to align with the Instruction Set Architecture (ISA) of the backend.

\begin{table}[H]
    \centering
    \begin{tabular}{c}
    \includegraphics[width=0.47\textwidth]{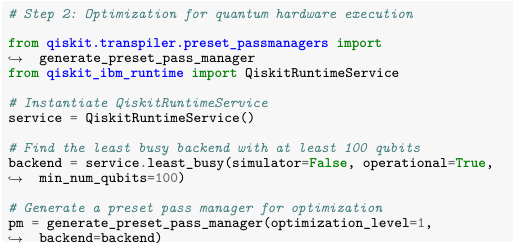} \\
    \includegraphics[width=0.47\textwidth]{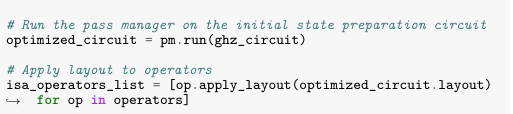} \\
    \end{tabular}
    \captionsetup{labelformat=empty, labelsep=none} 
\end{table}

$\bullet$ \textit{Execution on quantum hardware}. 
Proceed to submit the job for execution on the quantum hardware, implement error suppression using a technique known as dynamical decoupling to mitigate errors, and  adjust the resilience level to determine the degree of error resilience desired. Higher resilience levels yield more precise results but require longer processing times.

\begin{table}[H]
    \centering
    \begin{tabular}{c}
    \includegraphics[width=0.47\textwidth]{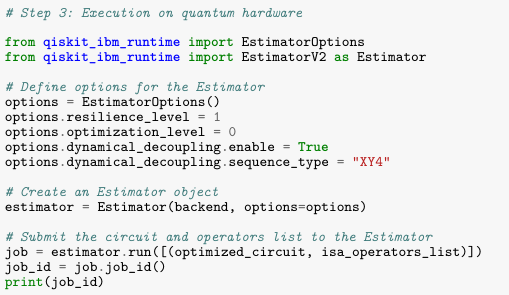} \\
    \end{tabular}
    \captionsetup{labelformat=empty, labelsep=none} 
\end{table}

$\bullet$ \textit{Post-processing of results}. Visualize the results through plotting after the job execution is finished. 
Observing the expectation value \( \langle Z_0 Z_i \rangle \) decreases as $i$ increases, indicating a deviation from the ideal scenario where all \( \langle Z_0 Z_i \rangle \) values should ideally be 1 in simulation.

\begin{table}[H]
    \centering
    \begin{tabular}{c}
    \includegraphics[width=0.47\textwidth]{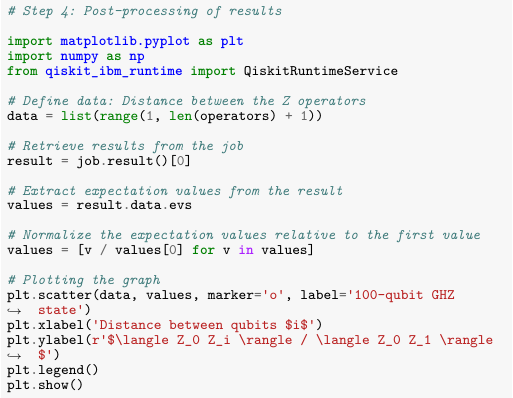} \\
    \end{tabular}
    \captionsetup{labelformat=empty, labelsep=none} 
\end{table}

\begin{figure}[H]
    \centering
    \includegraphics[width=0.45\textwidth]{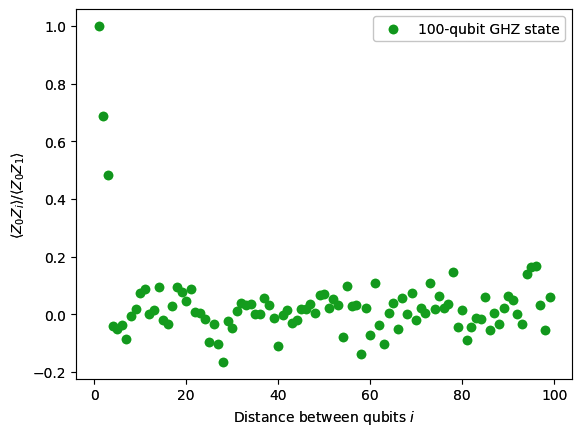}
\caption{Decay of normalized \( \langle Z_0 Z_i \rangle / \langle Z_0 Z_1 \rangle \) expectation values for a 100-qubit GHZ state, demonstrating quantum correlations as distance between qubits increases. Reproduced from~\citep{scale_qiskit}.}
    \label{fig:Z0ZI}
\end{figure}

For the 100-qubit GHZ state, the normalized expectation value \( \langle Z_0 Z_i \rangle / \langle Z_0 Z_1 \rangle \) is crucial for understanding the quantum correlations among the qubits. 
Figure~\ref{fig:Z0ZI} illustrates how the signal decays with increasing distance between qubits, reflecting the presence of noise in the system. To delve deeper into quantum computing with Qiskit, we refer to the recent review in~\citep{QCQiskit}.

%% file: 06Path.tex
\section{Toward Useful Quantum Computing} \label{S:Path}

\subsection{The IBM's era of quantum utility}

In 2023, an IBM and UC Berkeley groundbreaking experiment revealed a path toward practical quantum computing~\citep{IBM-Berkeley,IBM-Berkeley0}. This experiment demonstrated that quantum computers could execute circuits beyond the capabilities of brute-force classical simulations. For the first time, IBM Quantum has hardware and software capable of running quantum circuits at a scale of 100 qubits and 3,000 gates without prior knowledge of the outcomes~\citep{Utility_3,Gambetta_utility}. 

These advancements have prompted IBM Quantum to advocate moving beyond traditional circuit models by embracing parallelism, concurrent classical computing, and dynamic circuits. IBM Quantum emphasizes the necessity of a heterogeneous computing architecture that integrates scalable, parallel circuit execution with advanced classical computation~\citep{Utility_3,Gambetta_utility}.

IBM Quantum's vision for the future involves quantum-centric supercomputing~\citep{QCQiskit16,QCQiskit78}. At the IBM Quantum Summit 2023, significant updates were announced, bringing us closer to this goal, alongside an extended roadmap outlining IBM Quantum's journey toward quantum-centric supercomputing over the next decade. This will enable more advanced utility-scale work and provide a seamless development environment for IBM Quantum's users, potentially even before achieving fault tolerance\citep{Gambetta_utility,roadmap-utility,Utility_3,IBM-Berkeley,qLDPC_atoms,eror_qLDPC}

\begin{figure*}
    \centering
    \rotatebox[origin=c]{90}{\includegraphics[width=\textheight,height=\textwidth]{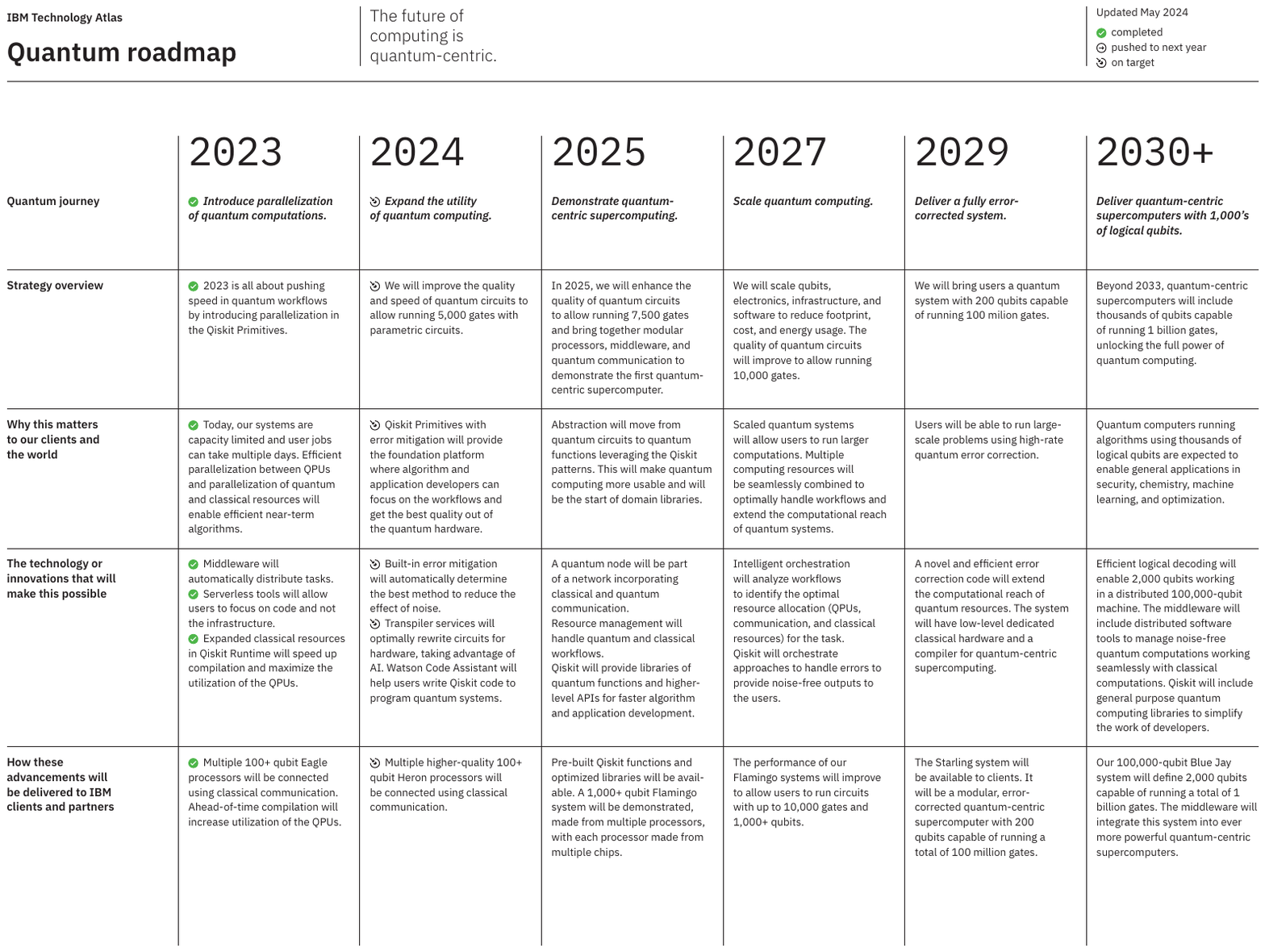}}
    \caption{IBM Quantum's technology roadmap, 
    starting with the introduction of parallelization of quantum computation in 2023,  
    aiming to expand the utility of quantum computing in 2024, and 
    advancing to the delivery of quantum-centric supercomputers with thousands  
    of logical qubits in 2030 and beyond. 
    Accessible at~\citep{Pdf_roadmap}. 
}
    \label{fig:roadmap1}
\end{figure*}

\subsection{IBM Quantum's roadmap}

To guide IBM Quantum's mission towards achieving quantum-centric supercomputing, IBM Quantum is extending its roadmap to 2033, spanning a decade of quantum innovation. The roadmap emphasizes advancements in the number of gates that IBM Quantum's processors and systems can execute. Beginning with \textit{Heron} targeting 5,000 gates by 2024, successive generations of processors will leverage improvements in quality to achieve increasingly higher gate counts~\citep{Gambetta_utility}.

In 2029, a pivotal milestone: the \textit{Starling} processor is projected to execute 100 million gates across 200 qubits, incorporating error correction based on the innovative Gross code. This code, represents an important advancement in error correction for near-term quantum computers~\citep{roadmap-utility,FullRoadmap-pdf}.

Following this, \textit{Blue Jay} is envisioned as a system capable of executing 1 billion gates across 2,000 qubits by 2033. This milestone signifies a nine-order-of-magnitude increase in gate execution capability since IBM Quantum first introduced its cloud-based devices in 2016~\citep{Gambetta_utility}.

The innovation roadmap will demonstrate the necessary technology to implement the Gross code through processors named \textit{Flamingo}, \textit{Crossbill}, and \textit{Kookaburra}, utilizing l-, m-, and c-couplers, respectively~\citep{roadmap-utility,FullRoadmap-pdf}.

\subsection{IBM Quantum safe}

Advancements in quantum technology highlight the need for new cryptographic methods based on mathematical challenges that are challenging for both quantum and classical computers to solve. IBM Quantum Safe aids enterprises in evaluating their cryptographic security and updating their cybersecurity strategies for the era of practical quantum computing.

IBM Quantum Safe roadmap outlines ongoing efforts to advance research in quantum-safe cryptography, collaborate with industry partners to promote adoption of post-quantum cryptographic solutions, and innovate new quantum-safe technologies. This includes ``IBM Quantum Safe Explorer," a cryptographic discovery tool launched in October 2023~\citep{qsafe,qsafeRep}.

Figure~\ref{fig:roadmap1} illustrates IBM's roadmap for quantum computing technology~\citep{Pdf_roadmap}. For 
detailed information on IBM Quantum's previous and updated development roadmaps, their accomplishments in hardware, software, execution, orchestration, and their innovation roadmap, we refer readers to~\citep{FullRoadmap-pdf}.

%% file: 07Conclusions.tex
\section{Conclusion}\label{S:Conclusion}

This study illuminates the rapid evolution and promising future of quantum computing within IBM Quantum. 
From pioneering quantum hardware advancements to robust software frameworks like Qiskit, IBM continues to redefine the boundaries of quantum technology.

We have delved into IBM Quantum’s dynamic journey in quantum computing, providing comprehensive performance evaluations of current and retired quantum computers that illustrate their evolution and future prospects. The documented metrics underscore IBM Quantum’s unwavering commitment to advancing quantum computing technology, offering a comparative framework across different systems and highlighting significant technological strides over time.

As we move towards quantum-centric supercomputing and quantum-safe cryptography, IBM Quantum’s trajectory promises transformative capabilities. With ongoing innovations and collaborative efforts, IBM Quantum is poised to ushering in a new era of computing prowess, offering profound implications across scientific, industrial, and cryptographic domains.

Looking ahead, the evolution of quantum computing will require addressing remaining challenges and exploring new avenues for innovation and application. 
Collaborative efforts within the quantum computing community, including partnerships and open-source initiatives, will be pivotal in unlocking the full potential of quantum technologies. Moreover, substantial funding and investments at national and international levels will be crucial in driving research, development, and deployment of quantum technologies.

%% file: 08Acknowledgment.tex
\begin{acknowledgments}

I am deeply grateful to the crowns of honor, though words cannot fully express my appreciation. The views and conclusions presented in this work reflect the author's perspective and do not necessarily align with those of IBM Quantum.

\end{acknowledgments}

\subsection*{Funding}

The author declares that no funding, grants, or other forms of support were received at any point throughout this research.

%% file: 10Appendix.tex
\newpage
\onecolumngrid 
\appendix

\section{Characteristics of IBM's Quantum Computers}\label{appendix:A1}

This section offers detailed performance summaries for various retired IBM Quantum's quantum computers, and some old performance data from current systems. Tables \ref{t:seattle} to \ref{t:lima} provide key specifications such as coherence times (T1 and T2), qubit frequencies, gate error rates, qubit counts, basis gates, connections, and calibration dates, documented for historical and educational purposes within the literature of NISQ computing era. 
This enables a comparative assessment of different IBM Quantum systems, highlighting advancements in technology over time. These retired systems, alongside quantum simulators, have laid the groundwork for newer generations that continue to push the boundaries of quantum computing.

\begin{table*}[htb]
    \centering
\caption{Hardware performance and qubit properties of the 433 qubit \textit{ibm\_seattle} quantum computer. The processor type is \textit{Osprey} r1. 
The basis gates of this machine are: ECR, ID, RZ, SX, X. With 
a median ECR error: $2.155 \times 10^{-2}$, 
a median SX error: $6.256\times 10^{-4}$, 
a median readout error: $4.910\times 10^{-2}$, 
a median T1: $88.35$ $\mu$s, and
a median T2: $58.73$ $\mu$s. 
Accessed June 29, 2023.}
\label{t:seattle}
    \begin{tabular}{c}
    \includegraphics[width=\textwidth]{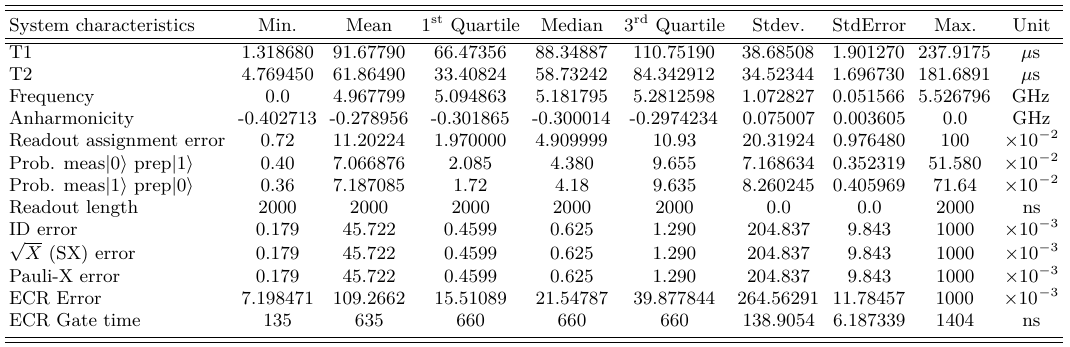} \\
    \end{tabular}
\end{table*}

\begin{table*}[htb]
    \centering
\caption{Summary of hardware performance, qubit characteristics and key specifications for the 127-qubits quantum computer \textit{ibm\_sherbrooke}, featuring basis gates ECR, ID, RZ, SX, and X. 
The processor type is \textit{Eagle} r3, and calibration data was accessed on June 30, 2023.}
\label{t:sherbrooke}
    \begin{tabular}{c}
    \includegraphics[width=\textwidth]{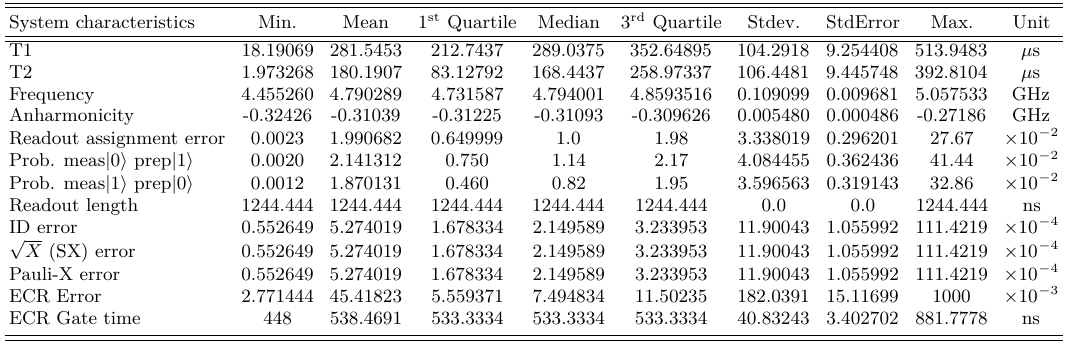} \\
    \end{tabular}
\end{table*}

\begin{table*}
    \centering
\caption{Summary of hardware performance, qubit characteristics and key specifications for the 127-qubits quantum computer \textit{ibm\_washington},  
featuring basis gates CX, ID, IF\_ELSE, RZ, SX and X. 
The processor type is \textit{Eagle} r3.  
The system boasts 288 CX qubit connections, and calibration data was accessed on November 12, 2022.}
\label{t:washington}
    \begin{tabular}{c}
    \includegraphics[width=\textwidth]{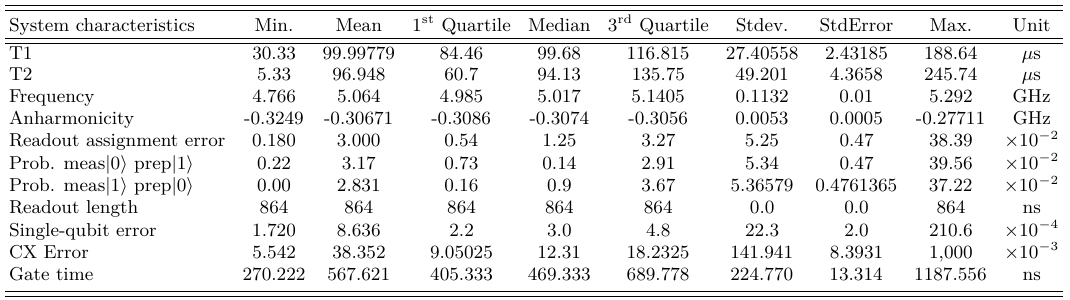} \\
    \end{tabular}
\end{table*}

\begin{table*}
    \centering
\caption{Summary of hardware performance, qubit characteristics and key specifications for the 127-qubit quantum computer \textit{ibm\_kyiv}, featuring basis gates CX, ID, RZ, SX and X.  The processor type is \textit{Eagle} r3.  The system boasts 288 CX qubit connections, and calibration data was accessed on November 14, 2022.\footnote{This system was upgraded, for the up-to-date specification see Table~\ref{tt:kyiv}.}}
\label{t:kyiv} 
    \begin{tabular}{c}
    \includegraphics[width=\textwidth]{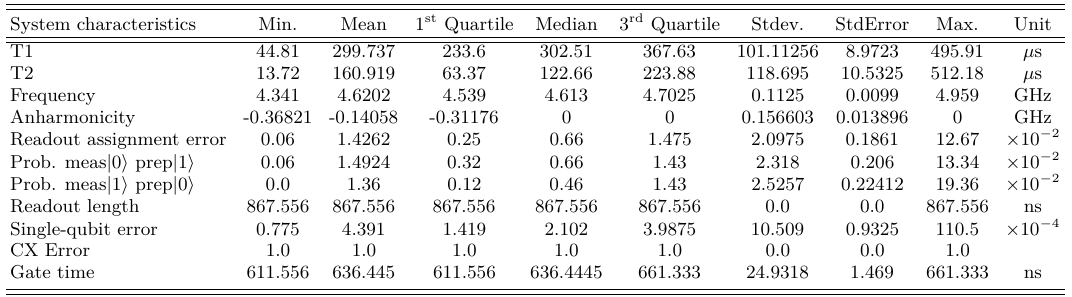} \\
    \end{tabular}
\end{table*}

\begin{table*}
    \centering
\caption{Summary of hardware performance, qubit characteristics and key specifications for the 65-qubit quantum computer \textit{ibm\_ithaca}, featuring basis gates CX, ID, RZ, SX and X. The processor type is \textit{Hummingbird} r3. 
The system boasts 144 CX qubit connections, and calibration data was accessed on November 14, 2022.}
\label{t:ithaca}
    \begin{tabular}{c}
    \includegraphics[width=\textwidth]{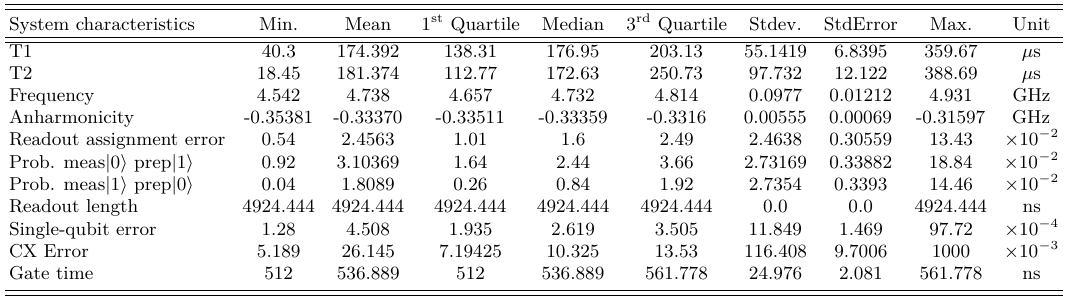} \\
    \end{tabular}
\end{table*}

\begin{table*}
    \centering
\caption{Summary of hardware performance, qubit characteristics and key specifications for the 27-qubit quantum computer \textit{ibmq\_kolkata}, featuring basis gates CX, ID, IF\_ELSE, RZ, SX and X.  The processor type is \textit{Falcon} r5.11. 
The system boasts 56 CX qubit connections, and calibration data was accessed on November 14, 2022.}
\label{t:kolkata}
    \begin{tabular}{c}
    \includegraphics[width=\textwidth]{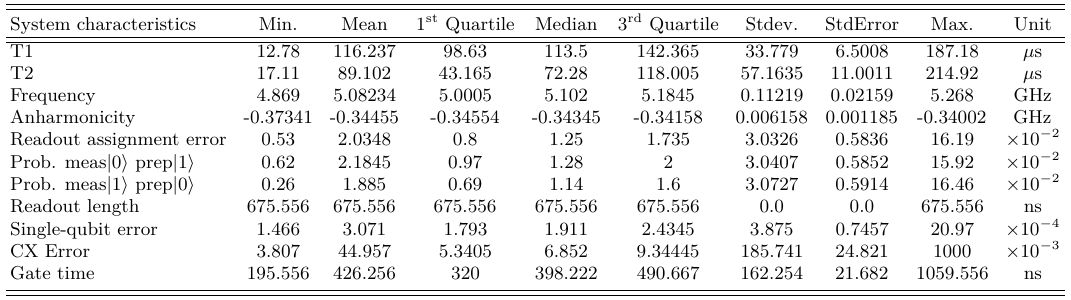} \\
    \end{tabular}
\end{table*}

\begin{table*}
    \centering
\caption{Summary of hardware performance, qubit characteristics and key specifications for the 27-qubit quantum computer \textit{ibmq\_montreal}, featuring basis gates CX, ID, RZ, SX and X. 
The processor type is \textit{Falcon} r4.  The system boasts 56 CX qubit connections, and calibration data was accessed on November 14, 2022.}
\label{t:montreal}
    \begin{tabular}{c}
    \includegraphics[width=\textwidth]{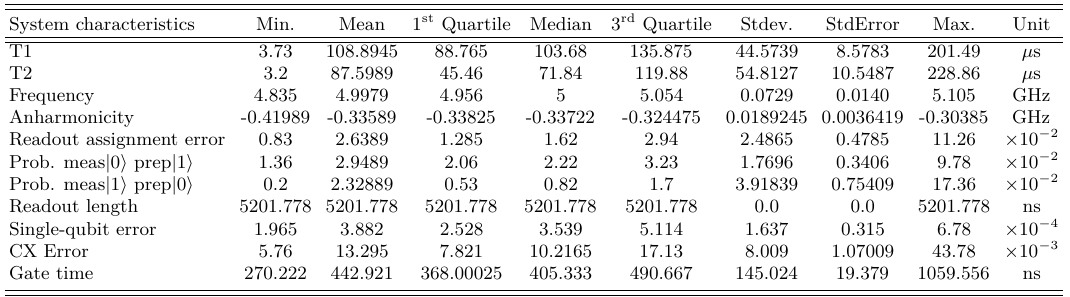} \\
    \end{tabular}
\end{table*}

\begin{table*}
    \centering
\caption{Summary of hardware performance, qubit characteristics and key specifications for the 27-qubit quantum computer \textit{ibmq\_mumbai}, featuring basis gates CX, ID, IF\_ELSE, RZ, SX and X. 
The processor type is \textit{Falcon} r5.10.  The system boasts 56 CX qubit connections, and calibration data was accessed on November 14, 2022.}
\label{t:mumbai}
    \begin{tabular}{c}
    \includegraphics[width=\textwidth]{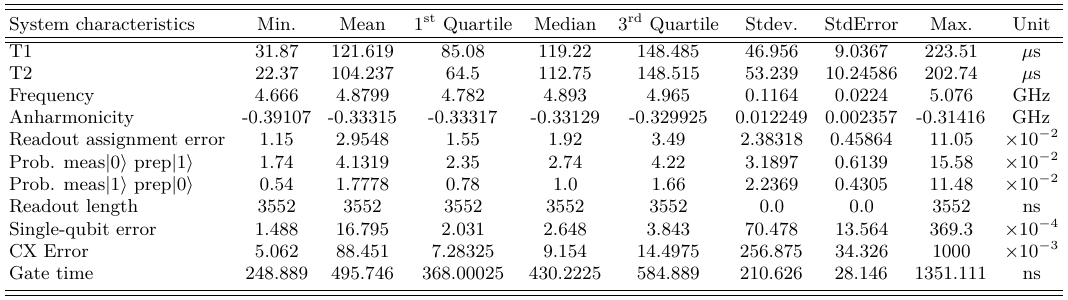} \\
    \end{tabular}
\end{table*}

\begin{table*}
    \centering
\caption{Summary of hardware performance, qubit characteristics and key specifications for the 27-qubit quantum computer \textit{ibm\_cairo}, featuring basis gates CX, ID, IF\_ELSE, RZ, SX and X. 
The processor type is \textit{Falcon} r5.11.  
The system boasts 56 CX qubit connections, and calibration data was accessed on November 14, 2022.}
\label{t:cairo}
    \begin{tabular}{c}
    \includegraphics[width=\textwidth]{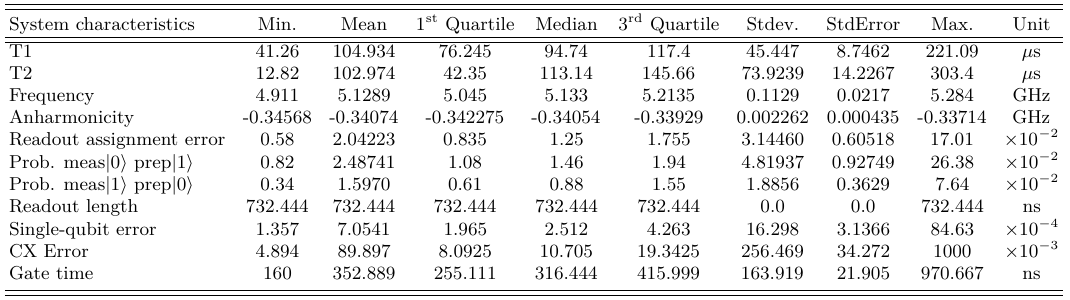} \\
    \end{tabular}
\end{table*}

\begin{table*}
    \centering
\caption{Summary of hardware performance, qubit characteristics and key specifications for the 27-qubit quantum computer \textit{ibm\_auckland}, featuring basis gates CX, ID, IF\_ELSE, RZ, SX and X.
The processor type is \textit{Falcon} r5.11.  The system boasts 56 CX qubit connections, and calibration data was accessed on November 14, 2022.}
\label{t:auckland}
    \begin{tabular}{c}
    \includegraphics[width=\textwidth]{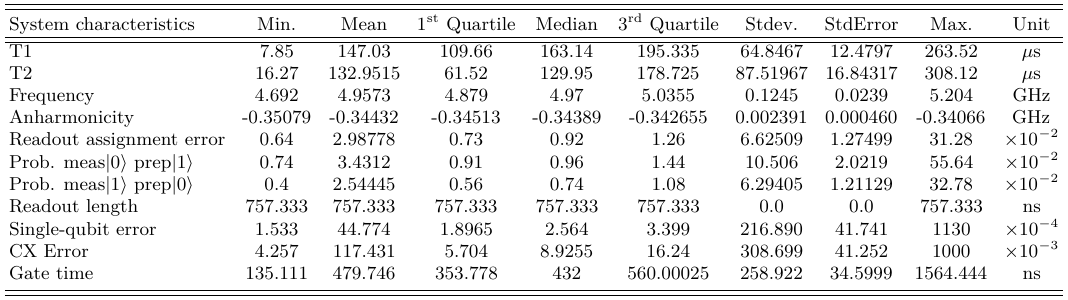} \\
    \end{tabular}
\end{table*}

\begin{table*}
    \centering
\caption{Summary of hardware performance, qubit characteristics and key specifications for the 27-qubit quantum computer \textit{ibm\_hanoi}, featuring basis gates CX, ID, IF\_ELSE, RZ, SX and X.  
The processor type is \textit{Falcon} r5.11.  The system boasts 56 CX qubit connections, and calibration data was accessed on November 14, 2022.}
\label{t:hanoi}
    \begin{tabular}{c}
    \includegraphics[width=\textwidth]{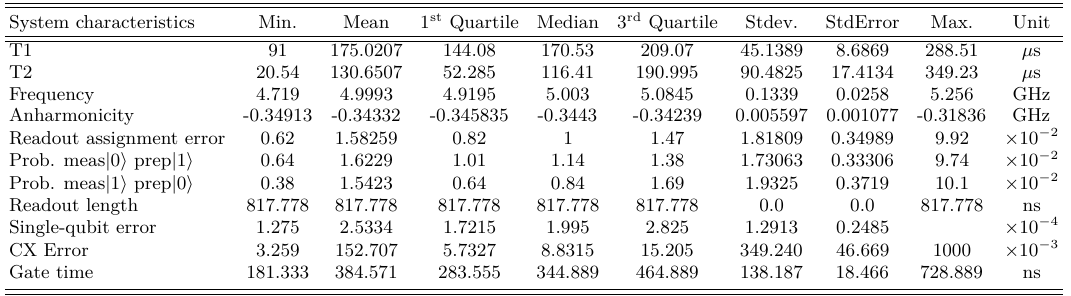} \\
    \end{tabular}
\end{table*}

\begin{table*}
    \centering
\caption{Summary of hardware performance, qubit characteristics and key specifications for the 27-qubit quantum computer \textit{ibm\_geneva}, featuring basis gates CX, ID, IF\_ELSE, RZ, SX and X.  The processor type is \textit{Falcon} r8.  The system boasts 56 CX qubit connections, and calibration data was accessed on November 14, 2022.}
\label{t:geneva}
    \begin{tabular}{c}
    \includegraphics[width=\textwidth]{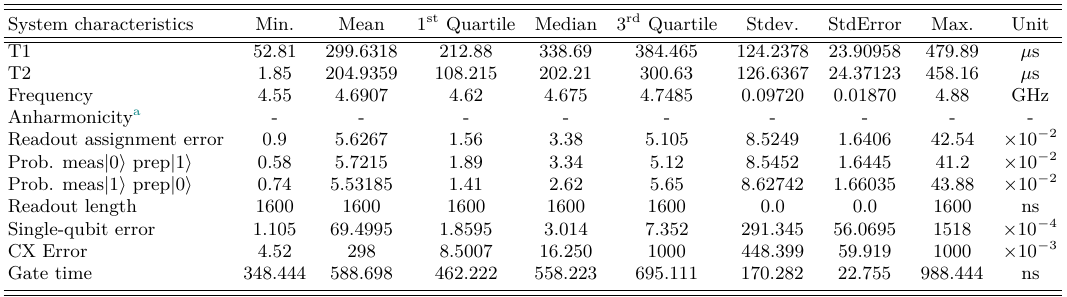} \\
    \end{tabular}
    \footnotetext[1]{Missing values.}
\end{table*}

\begin{table*}
    \centering
\caption{Summary of hardware performance, qubit characteristics and key specifications for the 27-qubit quantum computer \textit{ibmq\_toronto}, featuring basis gates CX, ID, RZ, SX and X.  The processor type is \textit{Falcon} r4.  The system boasts 56 CX qubit connections, and calibration data was accessed on November 14, 2022.}
\label{t:toronto}
    \begin{tabular}{c}
    \includegraphics[width=\textwidth]{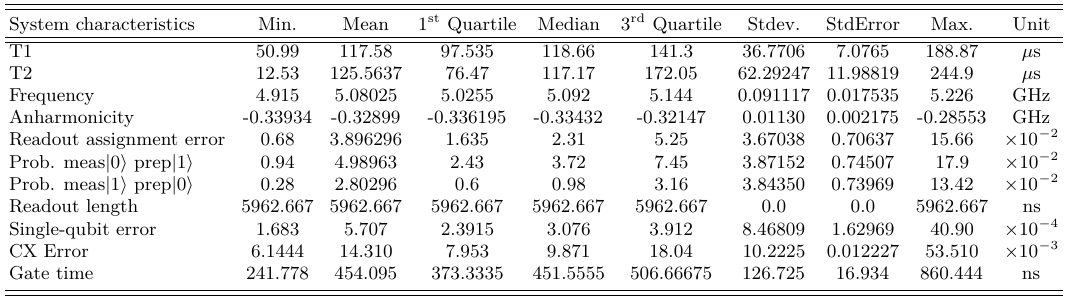} \\
    \end{tabular}
\end{table*}

\begin{table*}
    \centering
\caption{Summary of hardware performance, qubit characteristics and key specifications for the 27-qubit quantum computer \textit{ibm\_peekskill}, featuring basis gates CX, ID, IF\_ELSE, RZ, SX, X, X12, and XPRS.  The processor type is \textit{Falcon} r8.  
The system boasts 56 CX qubit connections, and calibration data was accessed on November 14, 2022.}
\label{t:peekskill}
    \begin{tabular}{c}
    \includegraphics[width=\textwidth]{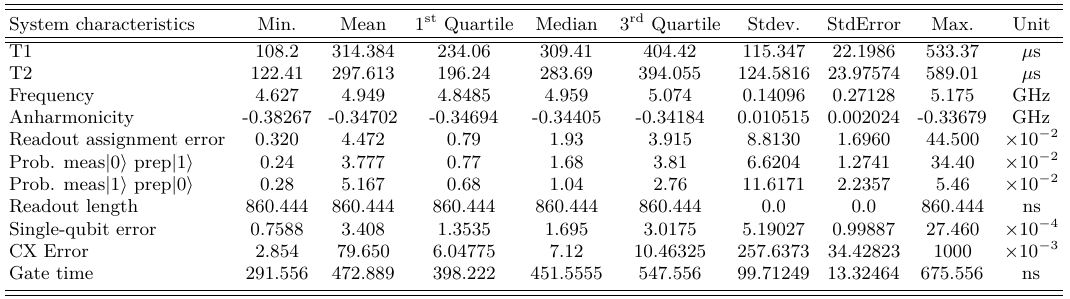} \\
    \end{tabular}
\end{table*}

\begin{table*}
    \centering
\caption{Summary of hardware performance, qubit characteristics and key specifications for the 16-qubit quantum computer \textit{ibmq\_guadalupe}, featuring basis gates CX, ID, RZ, SX and X. 
The processor type is \textit{Falcon} r4P.  The system boasts 32 CX qubit connections, and calibration data was accessed on November 14, 2022.}
\label{t:guadalupe}
    \begin{tabular}{c}
    \includegraphics[width=\textwidth]{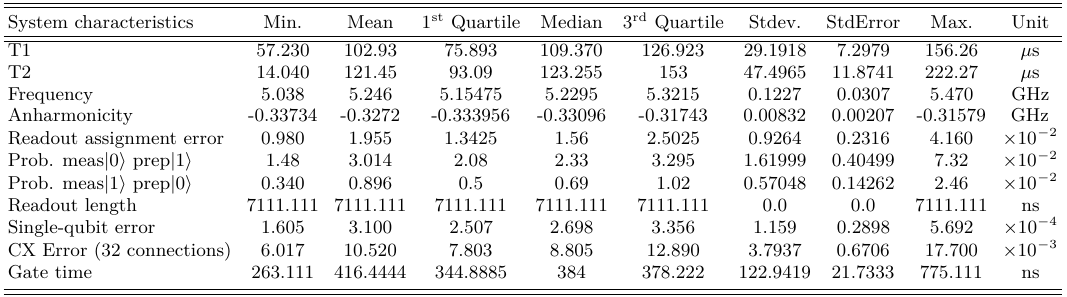} \\
    \end{tabular}
\end{table*}

\begin{table*}
    \centering
\caption{Summary of hardware performance, qubit characteristics and key specifications for the 7-qubit quantum computer \textit{ibm\_perth}, featuring basis gates CX, ID, IF\_ELSE, RZ, SX and X.  The processor type is \textit{Falcon} r5.11H.  The system boasts 12 CX qubit connections, and calibration data was accessed on November 13, 2022.}
\label{t:perth}
    \begin{tabular}{c}
    \includegraphics[width=\textwidth]{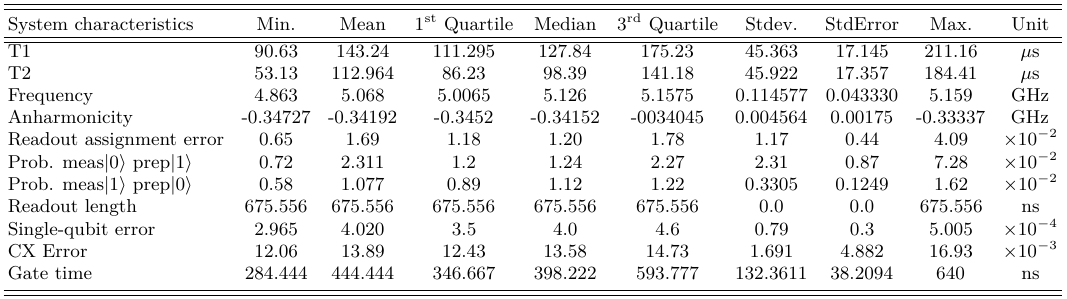} \\
    \end{tabular}
\end{table*}

\begin{table*}
    \centering
\caption{Summary of hardware performance, qubit characteristics and key specifications for the 7-qubit quantum computer \textit{ibm\_lagos}, featuring basis gates CX, ID, IF\_ELSE, RZ, SX and X.  The processor type is \textit{Falcon} r5.11H.  The system boasts 12 CX qubit connections, and calibration data was accessed on November 13, 2022.}
\label{t:lagos}
    \begin{tabular}{c}
    \includegraphics[width=\textwidth]{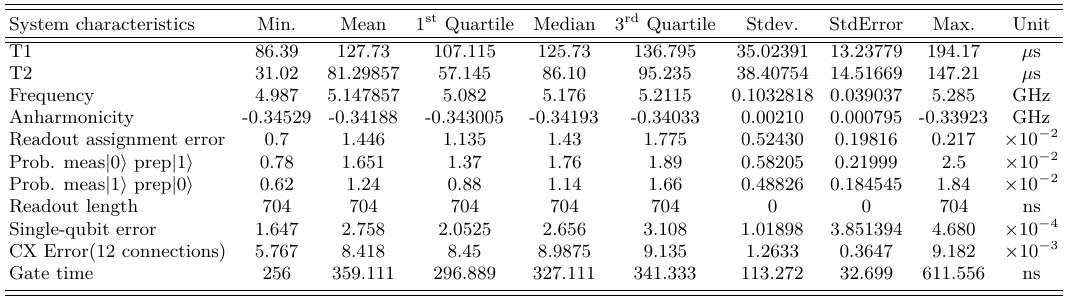} \\
    \end{tabular}
\end{table*}

\begin{table*}
    \centering
\caption{Summary of hardware performance, qubit characteristics and key specifications for the 7-qubit quantum computer \textit{ibm\_nairobi}, featuring basis gates CX, ID, IF\_ELSE, RZ, SX and X.  The processor type is \textit{Falcon} r5.11H.  The system boasts 12 CX qubit connections, and calibration data was accessed on November 13, 2022.}
\label{t:nairobi}
    \begin{tabular}{c}
    \includegraphics[width=\textwidth]{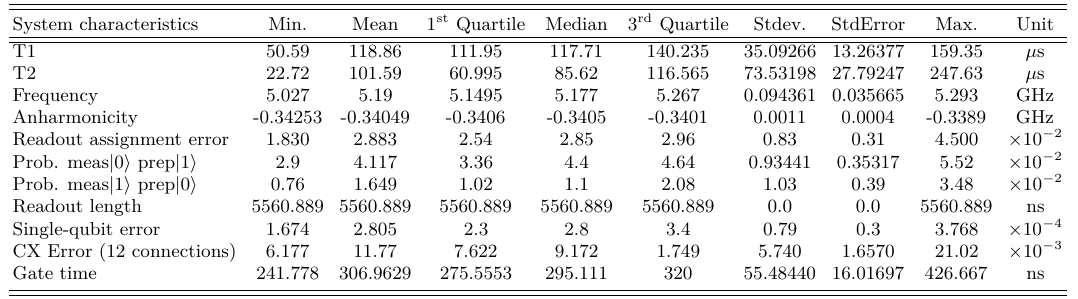} \\
    \end{tabular}
\end{table*}

\begin{table*}
    \centering
\caption{Summary of hardware performance, qubit characteristics and key specifications for the 7-qubit quantum computer \textit{ibm\_oslo}, featuring basis gates CX, ID, IF\_ELSE, RZ, SX and X.  The processor type is \textit{Falcon} r5.11H.  The system boasts 12 CX qubit connections, and calibration data was accessed on November 13, 2022.}
\label{t:oslo}
    \begin{tabular}{c}
    \includegraphics[width=\textwidth]{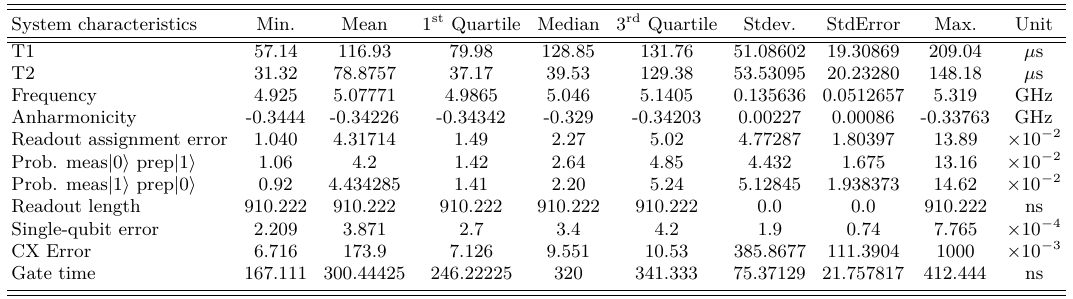} \\
    \end{tabular}
\end{table*}

\begin{table*}
    \centering
\caption{Summary of hardware performance, qubit characteristics and key specifications for the 7-qubit quantum computer \textit{ibmq\_jakarta}, featuring basis gates CX, ID, IF\_ELSE, RZ, SX and X.  The processor type is \textit{Falcon} r5.11H.  The system boasts 12 CX qubit connections, and calibration data was accessed on November 13, 2022.}
\label{t:jakarta}
    \begin{tabular}{c}
    \includegraphics[width=\textwidth]{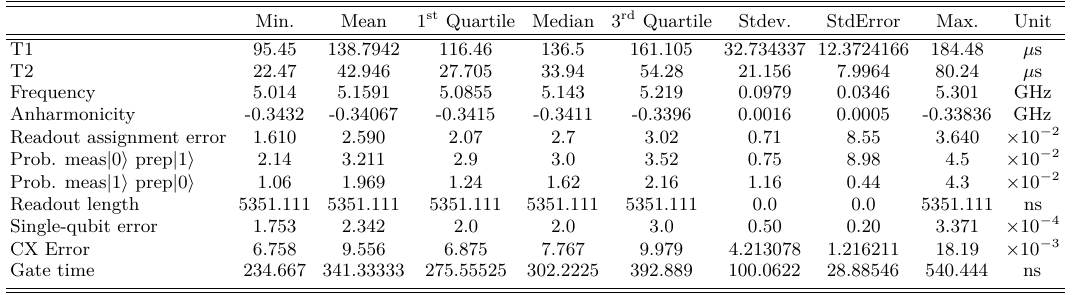} \\
    \end{tabular}
\end{table*}

\begin{table*}
    \centering
\caption{Summary of hardware performance, qubit characteristics and key specifications for the 5-qubit quantum computer \textit{ibmq\_manila}, featuring basis gates CX, ID, IF\_ELSE, RZ, SX, and X.  The processor type is \textit{Falcon} r5.11H.  The system boasts 8 CX qubit connections, and calibration data was accessed on November 13, 2022.}
\label{t:manila}
    \begin{tabular}{c}
    \includegraphics[width=\textwidth]{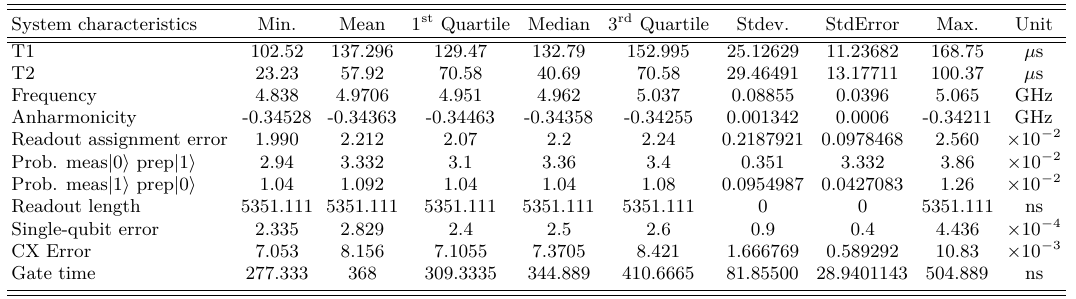} \\
    \end{tabular}
\end{table*}

\begin{table*}
    \centering
\caption{Summary of hardware performance, qubit characteristics and key specifications for the 5-qubit quantum computer \textit{ibmq\_quito}, featuring basis gates CX, ID, RZ, SX, and X.  The processor type is \textit{Falcon} r4T.  The system boasts 8 CX qubit connections, and calibration data was accessed on November 13, 2022.}
\label{t:quito}
    \begin{tabular}{c}
    \includegraphics[width=\textwidth]{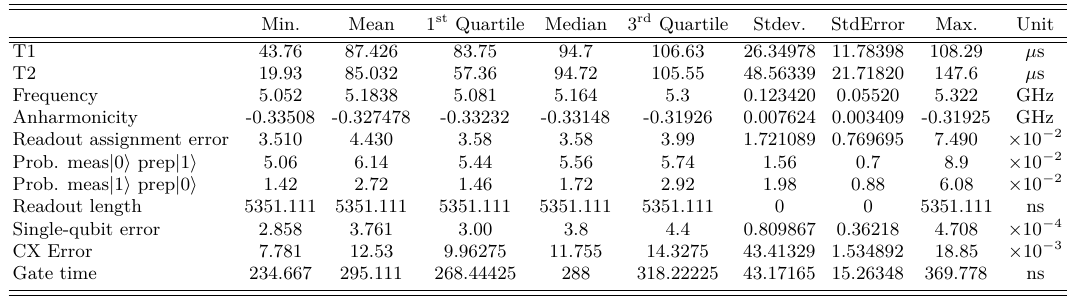} \\
    \end{tabular}
\end{table*}

\begin{table*}
    \centering
\caption{Summary of hardware performance, qubit characteristics and key specifications for the 5-qubit quantum computer \textit{ibmq\_belem}, featuring basis gates CX, ID, RZ, SX, and X.  The processor type is \textit{Falcon} r4T.  The system boasts 8 CX qubit connections, and calibration data was accessed on November 13, 2022.}
\label{t:belem}
    \begin{tabular}{c}
    \includegraphics[width=\textwidth]{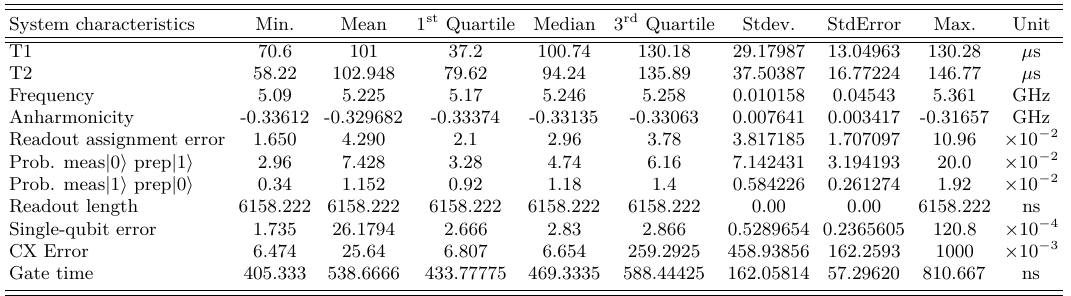} \\
    \end{tabular}
\end{table*}

\begin{table*}
    \centering
\caption{Summary of hardware performance, qubit characteristics and key specifications for the 5-qubit quantum computer \textit{ibmq\_lima}, featuring basis gates CX, ID, RZ, SX, and X.  The processor type is \textit{Falcon} r4T.  The system boasts 8 CX qubit connections, and calibration data was accessed on November 13, 2022.}
\label{t:lima}
    \begin{tabular}{c}
    \includegraphics[width=\textwidth]{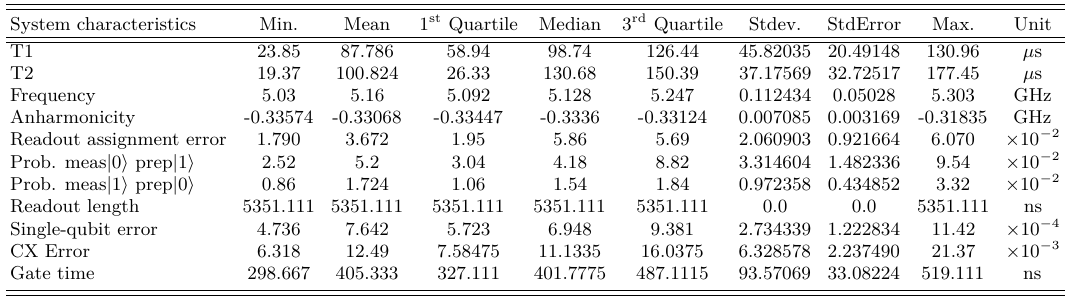} \\
    \end{tabular}
\end{table*}

%% file: IBMQuantum.bbl
\begin{thebibliography}{0}%
\makeatletter
\providecommand \@ifxundefined [1]{%
 \@ifx{#1\undefined}
}%
\providecommand \@ifnum [1]{%
 \ifnum #1\expandafter \@firstoftwo
 \else \expandafter \@secondoftwo
 \fi
}%
\providecommand \@ifx [1]{%
 \ifx #1\expandafter \@firstoftwo
 \else \expandafter \@secondoftwo
 \fi
}%
\providecommand \natexlab [1]{#1}%
\providecommand \enquote  [1]{``#1''}%
\providecommand \bibnamefont  [1]{#1}%
\providecommand \bibfnamefont [1]{#1}%
\providecommand \citenamefont [1]{#1}%
\providecommand \href@noop [0]{\@secondoftwo}%
\providecommand \href [0]{\begingroup \@sanitize@url \@href}%
\providecommand \@href[1]{\@@startlink{#1}\@@href}%
\providecommand \@@href[1]{\endgroup#1\@@endlink}%
\providecommand \@sanitize@url [0]{\catcode `\\12\catcode `\$12\catcode `\&12\catcode `\#12\catcode `\^12\catcode `\_12\catcode `\%12\relax}%
\providecommand \@@startlink[1]{}%
\providecommand \@@endlink[0]{}%
\providecommand \url  [0]{\begingroup\@sanitize@url \@url }%
\providecommand \@url [1]{\endgroup\@href {#1}{\urlprefix }}%
\providecommand \urlprefix  [0]{URL }%
\providecommand \Eprint [0]{\href }%
\providecommand \doibase [0]{https://doi.org/}%
\providecommand \selectlanguage [0]{\@gobble}%
\providecommand \bibinfo  [0]{\@secondoftwo}%
\providecommand \bibfield  [0]{\@secondoftwo}%
\providecommand \translation [1]{[#1]}%
\providecommand \BibitemOpen [0]{}%
\providecommand \bibitemStop [0]{}%
\providecommand \bibitemNoStop [0]{.\EOS\space}%
\providecommand \EOS [0]{\spacefactor3000\relax}%
\providecommand \BibitemShut  [1]{\csname bibitem#1\endcsname}%
\let\auto@bib@innerbib\@empty
\end{thebibliography}%


\begin{thebibliography}{9}

\bibitem{NISQ24} M. AbuGhanem and H. Eleuch, ``NISQ computers: a path to quantum supremacy," \textit{IEEE Access}, \textbf{12}, 102941-102961 (2024). 

\bibitem{principles} P. Dirac, ``The principles of quantum mechanics," Oxford, Clarendon Press, (1930).

\bibitem{AbuGMSc19} M. AbuGhanem, ``Properties of some quantum computing models,'' Master’s Thesis, Fac. Sci., Ain Shams Univ. (2019).

\bibitem{nisqQC11} D. R. Simon, ``On the power of quantum computation," \textit{SIAM J. Comput.} \textbf{26}(5), 1474–1483 (1997). 

\bibitem{nisqQC10} A. W. Harrow, A. Montanaro, ``Quantum computational supremacy," \textit{Nature} \textbf{549}(7671), 203–209 (2017). 

\bibitem{Light}  M. AbuGhanem, ``Information processing at the speed of light." Elsevier, \textit{SSRN} 4748781 (2024). http://dx.doi.org/10.2139/ssrn.4748781


\bibitem{PhotonicQuantumComputers} M. AbuGhanem, ``Photonic Quantum Computers," arXiv preprint arXiv:2409.08229, (2024). 


\bibitem{qsuperm1} J. Preskill, ``Quantum computing and the entanglement frontier," \textit{arXiv preprint} arXiv:1203.5813v3  (2012).

\bibitem{NISQ18} J. Preskill, ``Quantum Computing in the NISQ era and beyond," \textit{Quantum} \textbf{2}, 79 (2018).



\bibitem{IntoAsnG53} IBM Research. ``IBM Makes Quantum Computing Available on IBM Cloud to Accelerate Innovation." 
May 2016. \url{https://www.prnewswire.com/news-releases/ibm-makes-quantum-computing-available-on-ibm-cloud-to-accelerate-innovation-300262512.html}. 
Retrieved July 29 (2024). 

\bibitem{IntoAsnG55} W. J. Zeng. ``Unsupervised Machine Learning on Rigetti 19Q with Forest 1.2."
Dec. 18, 2017. \url{https://medium.com/rigetti/unsupervised-machine-learning-on-rigetti-19q-
with-forest-1-2-39021339699}. Retrieved July 29 (2024).

\bibitem{IntoAsnG91} Amazon. ``AWS Announces General Availability of Amazon Braket." August 13, 2020. \url{https://press.aboutamazon.com/2020/8/aws-announces-general-availability-of-amazon-braket}. Retrieved July 29 (2024).

\bibitem{IntoAsnG90} J. Russel. ``Honeywell Debuts Quantum System, ‘Subscription’ Business Model, and Glimpse of
Roadmap." Oct. 2020. \url{https://www.hpcwire.com/2020/10/29/honeywell-debuts-new-quantum-system-and-subscription-business-model/}. Retrieved July 29 (2024).

\bibitem{IntoAsnG93} Google Quantum AI. ``Google Quantum Computing Service." \url{https://quantumai.google/cirq/google/concepts}. Accessed July 29 (2024).


\bibitem{GoogleAI} M. AbuGhanem, Google Quantum AI’s Quest for Error-Corrected Quantum Computers. (2024).

\bibitem{IntoAsnG92} Xanadu AI. ``Xanadu launches photonic quantum cloud platform." September 2 (2020). 
\url{https://www.xanadu.ai/press/XANADU-LAUNCHES-PHOTONIC-QUANTUM-CLOUD-PLATFORM}. Retrieved July 29 (2024).

\bibitem{IntoAsnG94} Oxford Quantum Circuits. \url{https://oqc.tech/}. Accessed July 29 (2024). 

\bibitem{IntoAsnG96} H. Silverio. ``PASQAL - First Neutral Atoms Quantum Computer available on the cloud." May 6 (2022).
\url{https://www.pasqal.com/news/pasqal-first-neutral-atoms-quantum-computer-available-on-the-cloud/}. Retrieved July 29 (2024).

\bibitem{IntoAsnG95} QuEra. ``QuEra’s Quantum Computer ‘Aquila’ Now Available on Amazon Braket." November 1 (2022). 
\url{https://www.quera.com/press-releases/queras-quantum-computer-aquila-now-available-on-amazon-braket}. Retrieved July 29 (2024).

\bibitem{IntoAsnG97} Quandela. ``The first European quantum computer in the cloud, developed by Quandela." Nov. (2022). 
\url{https://www.quandela.com/wp-content/uploads/2022/11/Quandela-The-first-European-quantum-computer-on-the-cloud-developed-by-Quandela.pdf}. Accessed July 29 (2024).


\bibitem{IntoAsnG99} K. Svore. ``Azure Quantum is now in Public Preview." Feb. 2021. \url{https://cloudblogs.microsoft.com/quantum/2021/02/01/azure-quantum-preview/}. Retrieved July 29 (2024).

\bibitem{IntoAsnG98} Strangeworks. ``Strangeworks and IBM announce integration of IBM Quantum cloud services into the Strangeworks ecosystem." June 3, 2021. 
\url{https://strangeworks.com/press/strangeworks-and-ibm-announce-integration-of-ibm-quantum-cloud-services-into-the-strangeworks-ecosystem}. Retrieved July 29 (2024).

\bibitem{IntoAsnG100} Deutsch Telekom. ``T-Systems to offer quantum computing expertise and access to IBM Quantum computational resources." March 23, 2023. 
\url{https://www.telekom.com/en/media/media-information/archive/t-systems-launches-quantum-offering-1031464}. Retrieved July 29 (2024).

\bibitem{IntoAsnG101} Elsevier. ``Quantum computing research trends report." 
\url{https://www.elsevier.com/resources/quantum-computing-research-trends-report}. Accessed July 29 (2024).


\bibitem{Assessing} T. L. Scholten \textit{et al.} ``Assessing the benefits and risks of quantum computers," \textit{arXiv preprint} arXiv:2401.16317v2 (2024).


\bibitem{IBMQ} IBM Quantum, \url{https://quantum.ibm.com/}, accessed August (2024).

\bibitem{IBM-Berkeley} Y. Kim \textit{et al.} ``Evidence for the utility of quantum computing before fault tolerance." \textit{Nature} \textbf{618}, 500–505 (2023). 

\bibitem{Utility_3} G. Wendin and J. Bylander, ``Quantum computer scales up by mitigating errors," \textit{Nature} \textbf{618}, 462-463 (2023). 

\bibitem{QSimuIBM1} U. Alvarez-Rodriguez \textit{et al.} ``Quantum Artificial Life in an IBM Quantum Computer,'' \textit{Sci. Rep.} \textbf{8}, 14793 (2018).

\bibitem{QSimuIBM} IBM, ``Exploring quantum use cases for chemicals and petroleum: Changing how chemicals are designed and petroleum is refined,'' \url{https://www.ibm.com/downloads/cas/BDGQRXOZ}, accessed (2023).

\bibitem{QSimuIBM4} A. Kandala \textit{et al.} ``Hardware-efficient variational quantum eigensolver for small molecules and quantum magnets,'' \textit{Nature}, September 13 (2017). 

\bibitem{ExxonMobil} ExxonMobil, ``ExxonMobil and IBM to Advance Energy Sector Application of Quantum Computing,'' (January 8, 2019). \url{https://news.exxonmobil.com/press-release/exxonmobil-and-ibm-advance-energy-sector-application-quantum-computing}, accessed (2023).

\bibitem{roadmap-utility} IBM Debuts Next-Generation Quantum Processor \& IBM Quantum System Two, Extends Roadmap to Advance Era of Quantum Utility, December 4, 2023. \url{https://newsroom.ibm.com/2023-12-04-IBM-Debuts-Next-Generation-Quantum-Processor-IBM-Quantum-System-Two,-Extends-Roadmap-to-Advance-Era-of-Quantum-Utility}. Accessed July (2024).

\bibitem{Eagle’sperformance} O. Dial, ``Eagle’s quantum performance progress," 23 Mar 2022, \url{https://www.ibm.com/quantum/blog/eagle-quantum-processor-performance}

\bibitem{Gambetta_utility} J. Gambetta, ``The hardware and software for the era of quantum utility is here, We’ve entered a new era of quantum computing." December 4, 2023. \url{https://www.ibm.com/quantum/blog/quantum-roadmap-2033}, accessed July (2024).

\bibitem{1000-qubit} D. Castelvecchi, IBM releases first-ever 1,000-qubit quantum chip, ``The company announces its latest huge chip — but will now focus on developing smaller chips with a fresh approach to ‘error correction’." \textit{Nature} \textbf{624}, 238 (2023).

\bibitem{100-qubit} P. Ball, ``First quantum computer to pack 100 qubits enters crowded race," \textit{Nature} \textbf{599}, 542 (2021).

\bibitem{100qubits} J. Chow, O. Dial and J. Gambetta, ``IBM Quantum breaks the 100‑qubit processor barrier," 16 Nov. 2021 \url{https://www.ibm.com/quantum/blog/127-qubit-quantum-processor-eagle}, accessed (2024).

\bibitem{Ospprey} IBM Unveils 400 Qubit-Plus Quantum Processor and Next-Generation IBM Quantum System, \textit{IBM Newsroom}, Nov 9, 2022, \url{https://newsroom.ibm.com/2022-11-09-IBM-Unveils-400-Qubit-Plus-Quantum-Processor-and-Next-Generation-IBM-Quantum-System-Two}, accessed (2023).

\bibitem{ibm433} D. L. Underwood \textit{et al.} ``Using cryogenic CMOS control electronics to enable a two-qubit cross-resonance gate," \textit{arXiv:2302.11538}, (2023).

\bibitem{Qiskit} Qiskit, \url{https://www.ibm.com/quantum/qiskit}, accessed (2024).

\bibitem{Stef14} J. Chow \textit{et al.} ``Implementing a strand of a scalable fault-tolerant quantum computing fabric." \textit{Nat Commun.} \textbf{5}, 4015 (2014).


\bibitem{Processortypes} Processor types, IBM Quantum documentation, \url{https://docs.quantum.ibm.com/guides/processor-types}, accessed July (2024).


\bibitem{cross-resonance} S. Sheldon, E. Magesan and J. Chow, J. Gambetta,  ``Procedure for systematically tuning up cross-talk in the cross-resonance gate." \textit{Phys. Rev. A} \textbf{93}, 060302(R). (2016).


\bibitem{QPU}  Quantum processing units, IBM Quantum platform, \url{https://quantum.ibm.com/services/resources}, accessed July (2024).


\bibitem{QCQiskit} A. Javadi-Abhari \textit{et al.} ``Quantum computing with Qiskit,"  \textit{arXiv preprint} arXiv:2405.08810 (2024).


\bibitem{QCQiskit10} The State of quantum open source software (2023): survey results. \url{https://unitary.fund/posts/2023_survey_results}.


\bibitem{QCQiskit42} R. S. Gupta \textit{et al.} ``Encoding a magic state with beyond break-even fidelity." arXiv
preprint arXiv:2305.13581 (2023).


\bibitem{Entangling} M. AbuGhanem and H. Eleuch, ``Two-qubit entangling gates for superconducting quantum computers," \textit{Results in Physics} \textbf{56}, 107236 (2024). 

\bibitem{QCQiskit21} E. B{\"a}umer \textit{et al.} ``Efficient Long-Range Entanglement using Dynamic Circuits." \textit{arXiv preprint} arXiv:2308.13065 (2023).

\bibitem{QCQiskit40} R. C. Farrell \textit{et al.} ``Quantum Simulations of Hadron Dynamics in the Schwinger Model using 112 Qubits." \textit{arXiv preprint} arXiv:2401.08044 (2024). 

\bibitem{SQSCZ1} M. AbuGhanem \textit{et al.} ``Fast universal entangling gate for superconducting quantum computers," \textit{Elsevier}, \textit{SSRN}, 4726035 (2024). http://dx.doi.org/10.2139/ssrn.4726035

\bibitem{QCQiskit30} E. H. Chen \textit{et al.} ``Realizing the Nishimori transition across the error threshold for constant-depth quantum circuits." \textit{arXiv preprint} arXiv:2309.02863 (2023). 

\bibitem{QCQiskit39} R. C. Farrell \textit{et al.} ``Scalable circuits for preparing ground states on digital quantum computers: The schwinger model vacuum on 100 qubits." \textit{PRX Quantum} \textbf{5}:020315 (2024). 

\bibitem{QCQiskit59} R. Majumdar \textit{et al.} ``Best practices for quantum error mitigation with digital zero-noise extrapolation." \textit{arXiv preprint} arXiv:2307.05203 (2023). 

\bibitem{SQSCZ2} M. AbuGhanem, ``Full quantum process tomography of a universal entangling gate on an IBM's quantum computer,"  \textit{arXiv preprint} arXiv:2402.06946 (2024).

\bibitem{QCQiskit65} J. A. Montanez-Barrera and K. Michielsen. ``Towards a universal QAOA protocol: Evidence of quantum advantage in solving combinatorial optimization problems." \textit{arXiv preprint} arXiv:2405.09169 (2024). 


\bibitem{QCQiskit72} E. Pelofske \textit{et al.} ``Scaling Whole-Chip QAOA for Higher-Order Ising Spin Glass Models on Heavy-Hex Graphs." \textit{arXiv preprint} arXiv:2312.00997 (2023).

\bibitem{GSA} M. AbuGhanem, ``Comprehensive characterization of three-qubit Grover search algorithm on IBM's 127-qubit superconducting quantum computers," \textit{arXiv preprint} arXiv:2406.16018 (2024).


\bibitem{QCQiskit101} H. Yu \textit{et al.} ``Simulating large-size quantum spin chains on cloud-based superconducting quantum computers." \textit{Phys. Rev. Res.}, 5(1):013183 (2023).

\bibitem{QCQiskit100} T. Yasuda \textit{et al.} ``Quantum reservoir computing with repeated measurements on superconducting devices." \textit{arXiv preprint} arXiv:2310.06706 (2023). 


\bibitem{EPJQ} M. AbuGhanem and H. Eleuch, ``Full quantum tomography study of Google's Sycamore gate on IBM’s quantum computers," \textit{EPJ Quantum Technology} \textbf{11}(1), 36 (2024).

\bibitem{Googles} M. AbuGhanem and H. Eleuch, ``Experimental characterization of Google’s Sycamore quantum AI on an IBM’s quantum computer,"  \textit{Elsevier}, \textit{SSRN} 4299338  (2023). \url{http://dx.doi.org/10.2139/ssrn.4299338}. 


\bibitem{QCQiskit104} V. Zhang and P. D. Nation. ``Characterizing quantum processors using discrete time crystals." \textit{arXiv preprint} arXiv:2301.07625 (2023). 

\bibitem{QCQiskit29} M. Cerezo \textit{et al.} ``Variational quantum algorithms." \textit{Nature Reviews Physics}, \textbf{3}(9):625–644 (2021).


\bibitem{QCQiskit16} Y. Alexeev \textit{et al.} ``Quantum-centric supercomputing for materials science: A perspective on challenges and future directions." Future Generation Computer Systems, (2024).

\bibitem{QCQiskit78} J. Robledo-Moreno \textit{et al.} ``Chemistry Beyond Exact Solutions on a Quantum-Centric Supercomputer." \textit{arXiv preprint} arXiv:2405.05068 (2024). 

\bibitem{QCQiskit38} I. Faro \textit{et al.} ``Middleware for quantum: An orchestration of hybrid quantum-classical systems." \textit{In 2023 IEEE International Conference on Quantum Software (QSW)} 1-8. IEEE (2023).


\bibitem{scale_qiskit} Scale to large number of qubits, IBM Quantum documentation, 
\url{https://docs.quantum.ibm.com/guides/hello-world#scale-to-large-numbers-of-qubits}, accessed July (2024).


\bibitem{QCQiskit31} A. Cross \textit{et al.} ``OpenQASM 3: A broader and deeper quantum assembly language." \textit{ACM Trans. Quan. Comp.}, 3(3):1–50 (2022).

\bibitem{QCQiskit54} D. Kremer \textit{et al.} ``Practical and efficient quantum circuit synthesis and transpiling with reinforcement learning." \textit{arXiv preprint} arXiv:2405.13196 (2024).


\bibitem{QCQiskit53} A. Yu. Kitaev. ``Quantum measurements and the Abelian stabilizer problem." \textit{arXiv preprint}  arXiv:quant-ph/9511026,(1995).


\bibitem{QCQiskit23} C. H. Bennett, D. P. DiVincenzo, J. A. Smolin, and W. K. Wootters. ``Mixed-state entanglement and quantum error correction." 
\textit{Phys. Rev. A}, \textbf{54}(5):3824 (1996).

\bibitem{QCQiskit58} S. Lloyd. ``Universal quantum simulators." \textit{Science}, \textbf{273}(5278):1073–1078 (1996).

\bibitem{QCQiskit22} C. H. Bennett \textit{et al.} 
``Purification of noisy entanglement and faithful teleportation via noisy channels." \textit{Phys. rev. lett.}, \textbf{76}(5):722 (1996).

\bibitem{QCQiskit95} L. Viola, E. Knill, and S. Lloyd. ``Dynamical decoupling of open quantum systems." \textit{Phys. Rev. Lett.}, \textbf{82}(12):2417 (1999).

\bibitem{IBM-Berkeley0} R. Mandelbaum, ``A new paper from IBM and UC Berkeley shows a path toward useful quantum computing. A useful application for 127-qubit quantum processors with error mitigation." June 14, 2023. \url{https://www.ibm.com/quantum/blog/utility-toward-useful-quantum}



\bibitem{qLDPC_atoms} Q. Xu \textit{et al.} ``Constant-Overhead Fault-Tolerant Quantum Computation with Reconfigurable Atom Arrays," \textit{arXiv preprint}  arXiv:2308.08648v1  (2023). 


\bibitem{eror_qLDPC} S. Bravyi \textit{et al.} ``High-threshold and low-overhead fault-tolerant quantum memory," \textit{Nature} \textbf{627}, 778-782 (2024).



\bibitem{FullRoadmap-pdf} Development \& Innovation Roadmap, IBM Quantum, \url{https://www.ibm.com/quantum/assets/IBM_Quantum_Development_&_Innovation_Roadmap.pdf}, accessed August (2024).

\bibitem{Pdf_roadmap} IBM Quantum's technology roadmap, \url{https://www.ibm.com/roadmaps/quantum.pdf}, Accessed August (2024). 


\bibitem{qsafe} Make the world quantum safe, \url{https://www.ibm.com/quantum/quantum-safe}, Accessed August 1st (2024).

\bibitem{qsafeRep} IBM Quantum Safe Explorer: Simplify the discovery of cryptography and the management of quantum security risks, IBM Quantum, 
\url{https://www.ibm.com/downloads/cas/O5B0WXVZ}. Accessed August 1st (2024).











\end{thebibliography}
